\documentclass[useAMS,usenatbib]{mn2e}
\usepackage{hyperref}
\usepackage{amssymb,esint}
\usepackage{graphicx}
\usepackage{epstopdf}
\usepackage{mathptmx}
\newcommand{\bea}{\begin{eqnarray}}
\newcommand{\eea}{\end{eqnarray}}
\newcommand{\beq}{\begin{equation}}
\newcommand{\eeq}{\end{equation}}

\title[Superstructures in the SDSS DR7 surveys]{A robust public catalogue of voids and superclusters in the SDSS Data Release 7 galaxy surveys}
\author[S. Nadathur \& S. Hotchkiss]{Seshadri Nadathur$^{1,2}$ \& Shaun Hotchkiss$^{2,3}$\\
$^1$Fakult\"at f\"ur Physik, Universit\"at Bielefeld, Postfach 100131, D-33501 Bielefeld, Germany\\
$^2$Department of Physics, University of Helsinki and Helsinki Institute of Physics, P.O. Box 64, FIN-00014, University of Helsinki, Finland\\
$^3$Department of Physics and Astronomy, University of Sussex, Brighton, BN1 9QH, UK}

\begin{document}

\date{\today}

\pagerange{\pageref{firstpage}--\pageref{lastpage}}

\label{firstpage}

\maketitle

\begin{abstract}
The study of the interesting cosmological properties of voids in the Universe depends on the efficient and robust identification of such voids in galaxy redshift surveys. Recently, \cite{Sutter:2012wh} have published a public catalogue of voids in the Sloan Digital Sky Survey Data Release 7 main galaxy and luminous red galaxy samples, using the void-finding algorithm {\small ZOBOV}, which is based on the watershed transform. We examine the properties of this catalogue and show that it suffers from several problems and inconsistencies, including the identification of some extremely overdense regions as voids. As a result, cosmological results obtained using this catalogue need to be reconsidered. We provide instead an alternative, self-consistent, public catalogue of voids in the same galaxy data, obtained from using an improved version of the same watershed transform algorithm. We provide a more robust method of dealing with survey boundaries and masks, as well as with a radially varying selection function, which means that our method can be applied to any other survey. We discuss some basic properties of the voids thus discovered, and describe how further information may be obtained from the catalogue. In addition, we apply an inversion of the algorithm to the same data to obtain a corresponding catalogue of large-scale overdense structures, or ``superclusters". Our catalogues are available for public download at \url{www.hip.fi/nadathur/download/dr7catalogue}.
\end{abstract}

\label{firstpage}

\maketitle

\begin{keywords}
catalogues -- cosmology: observations -- surveys -- large-scale structure of Universe
\end{keywords}


\section{Introduction}
\label{section:intro}

A ubiquitous feature of galaxy redshift surveys are large underdense regions of space, called voids, that are only sparsely populated with galaxies. While most of the galaxies are concentrated in dense clusters and filaments, voids comprise most of the volume of the Universe. Since the discovery of the first voids (\citealt*{Gregory:1978,Joeveer:1978}; \citealt{Kirshner:1981wz}), they have been recognised as important objects for study whose properties may test models of structure formation. 

The importance of voids derives from the fact that they are largely empty of matter and therefore their dynamics are dominated by dark energy from early times \citep*{Goldberg:2004}. This affects the behaviour of gravitational clustering and galaxy formation within the void, which may affect the properties of galaxies within voids \citep[see][and references within]{Pan:2011hx}. Voids are also interesting objects for other cosmological studies. The Alcock-Paczynski test \citep{Alcock:1979} can be applied to the  observed shapes of voids in redshift space, which allows a reconstruction of the expansion history of the Universe \citep{Ryden:1995,Ryden:1996,Biswas:2010,Lavaux:2009wm}. Void alignments, spins and ellipticities have also been suggested as cosmological probes \citep{Lee:2006gj,Park:2007,Lee:2007kq}. Comparisons between low-redshift and high-redshift voids may test the $\Lambda$CDM model \citep*{Viel:2008}.The abundance of voids is sensitive to any non-Gaussianity in the primordial density perturbations (\citealt*{Kamionkowski:2009}; \citealt{D'Amico:2010kh}). As this abundance is also dependent on the growth rate of structure it may be used to test models of modified gravity \citep*{Li:2009,Li:2011pj}.

Voids themselves also have measurable gravitational effects, which can be extracted in combination with other data sets. The gravitational lensing signal of stacked voids may be obtained from galaxy imaging surveys \citep{Krause:2013}. \citet*{Granett:2008ju} obtained a high-significance detection of the imprint of voids on the cosmic microwave background (CMB) via the integrated Sachs-Wolfe (ISW) effect, though the measured amplitude of this signal appears to be unusually large \citep*{Hunt:2008wp,Nadathur:2011iu,Flender:2012wu,HernandezMonteagudo:2012ms}.

All such observational studies of voids require large, reliable and self-consistently identified catalogues of observed voids that cover a wide range of redshifts. For this purpose one must use a void finding algorithm applied to galaxy survey data. There are many different void finding algorithms, which operate on different principles and may produce somewhat different results \citep[see][for a summary and comparisons]{Colberg:2008}. The construction of some previous catalogues has used a method based on overlapping spheres of underdensities \citep{Hoyle:2003hc,Pan:2011hx}; however, this method uses some rather fine-tuned parameters, imposes assumptions about void shapes, and ignores the presence of survey masks and boundaries, so may be regarded as less than ideal. In addition, these catalogues only extend to rather low redshifts ($z\sim0.1$), which restricts the use of these voids for several of the cosmological tests described above. With the exception of the assumption about void shapes, similar concerns apply to the catalogue of \citet*{Tavasoli:2013}.

Recently,  \citet{Sutter:2012wh} presented a catalogue of voids obtained from the use of a modified version of the watershed-transform void finding algorithm {\small ZOBOV} \citep{Neyrinck:2007gy} applied to data from the Sloan Digital Sky Survey (SDSS) Data Release 7 \citep{Abazajian:2008wr}. This catalogue extends to redshifts $z\gtrsim0.4$. The parameter-free {\small ZOBOV} void finder makes no assumptions about the shapes of voids, and uses only topological information about the density field of galaxies, which is reconstructed using a Voronoi tessellation of the survey data. {\small ZOBOV} also provides natural ways to account for survey boundaries and for variable selection functions across the survey region, which means that the voids thus found should in principle be better suited for use in cosmology than those from other void-finders lacking these advantages. Indeed, this catalogue of voids has been used in a variety of recent studies, e.g. \citet{Sutter:2012tf,Ilic:2013cn,Planck:ISW,Pisani:2013yxa,Hamaus:2013,Melchior:2013,Sutter:2013ssy}.

Unfortunately, however, this catalogue suffers from several inconsistencies which have not previously been recognized. These are due partly to the choice of inappropriate criteria to govern the operation of the void finding algorithm, and partly to inconsistent application of even these criteria. In addition, we identify a high level of contamination from survey boundary effects. We discuss the various problems with this catalogue in detail, and demonstrate that they result in the inclusion within the catalogue of many overdense structures wrongly classified as voids. This means that the identified void locations do not, on average, correspond to underdense regions in the galaxy density field. This lack of self-consistency of the catalogue has direct implications for results of the studies mentioned above, which may need to be reconsidered.

Nevertheless, a catalogue of voids from the SDSS DR7 data is clearly a desirable product. We also feel that the advantages of the watershed void-finding approach mean that it is well suited to the task of identifying such voids. We therefore present an alternative, self-consistent, public catalogue of voids found in the same volume-limited SDSS galaxy surveys using our own modification of the {\small ZOBOV} algorithm after correcting the problems with the \citet{Sutter:2012wh} methodology. We introduce stricter controls to limit the artificial effects of the survey mask, and apply a correction for the radial variation of the mean galaxy number density. 

We also investigate in more detail the operation of {\small ZOBOV} when applied to galaxy data, and discuss the choice of void selection criteria. This choice results in a small and unavoidable degree of subjectivity to the final result, and the most appropriate choice may vary depending on the intended use of the catalogue. We present results for two restricted subsets of the catalogue based on choices that we believe are well-motivated, but we also make available the more general catalogue so that users may apply their own selection criteria according to need. We compare the list of voids we obtain to those in the existing catalogue and show that our algorithm performs better at correctly identifying locations of underdensities.

We then invert the void-finding algorithm to provide the first public catalogue of ``superclusters" in the same galaxy data. These superclusters are large-scale overdense structures in the galaxy distribution, but are not necessarily gravitationally bound objects. They correspond to large-scale peaks in the density field in the same way that voids correspond to large-scale troughs. Previous studies of superclusters have used different approaches to identifying them. Some \citep[e.g.][]{Einasto:1997zd,Berlind:2006} employ a friends-of-friends (FoF) method, but this is very sensitive to the arbitrary choice of linking length, and could give rise to surprising conclusions if not carefully used \citep{Park:2012dn,Nadathur:2013mva}. Another approach is to use smoothing kernels on the density field \citep[e.g.][]{Einasto:2003us,Einasto:2006mm,Einasto:2011zc, Liivamagi:2012}, but this imposes a spherical symmetry that is probably artificial. The watershed algorithm avoids both of these problems by capturing the full geometry of the superclusters from topological information of the density field.

We expect the superclusters we find to be cosmologically interesting for the many of the same reasons as voids. In particular, their gravitational effects may be measured by lensing and ISW studies. Indeed the supercluster structures presented in our catalogue are in principle most similar to those found in the SDSS DR6 photometric catalogue of luminous red galaxies \citep{AdelmanMcCarthy:2007aa}, which have already been used for ISW measurements \citet{Granett:2008ju,Granett:2008dz}. Due to the different redshift ranges of the galaxy samples we use, however, the two catalogues are essentially independent of each other.

The layout of the paper is as follows. We begin by describing the data samples we use in Section~\ref{section:data}. In Section~\ref{section:ZOBOV} we provide a brief description of the general operation of the {\small ZOBOV} void-finding algorithm. In Section~\ref{section:Sutter} we examine the previous public void catalogue of \citet{Sutter:2012wh} in detail, highlighting its inconsistencies and possible improvements to the algorithm. We then discuss the implementation of these improvements in our own modified void-finding algorithm in Section~\ref{section:ourcatalogue}. In Section~\ref{section:results} we present our results, including number counts of structures, their redshift and size distributions, and stacked radial density profiles. We conclude in Section~\ref{section:conclusions} with a discussion of the implication of our results and the future applications of our catalogue.

\section{Data samples}
\label{section:data}

In this paper, we use galaxy samples taken from the SDSS main galaxy redshift survey \citep{Strauss:2002dj} and the SDSS luminous red galaxy (LRG) survey \citep{Eisenstein:2001cq}. The data samples we use are identical to those used by \citet{Sutter:2012wh}; nevertheless, we briefly describe them again below.

The main galaxy sample we use is taken from the New York University Value-Added Galaxy Catalog \citep[NYU-VAGC;][]{Blanton:2004aa}, which is a catalogue of low-redshift ($z\lesssim0.3$) galaxies based on publicly-released surveys matched to galaxies from the SDSS \citep{Abazajian:2008wr} using improved photometric calibrations \citep{Padmanabhan:2007zd}. The absolute magnitudes $M_r$ are computed after applying evolution and $K$-corrections assuming a cosmology consistent with the WMAP 7-year results \citep{Komatsu:2010fb}, i.e. with $\Omega_M = 0.27$, $\Omega_\Lambda=0.73$ and $h=0.71$. 

Based on these magnitudes, we construct four uniform subsamples of the main galaxy catalogue, labelled \emph{dim1} ($M_r<-18.9$), \emph{dim2} ($M_r<-20.4$), \emph{bright1} ($M_r<-21.35$) and \emph{bright2} ($M_r<-22.05$). These can be chosen to span non-overlapping redshift bins, with $0<z<0.05$ for \emph{dim1}, $0.05<z<0.10$ for \emph{dim2}, $0.10<z<0.15$ for \emph{bright1} and $0.15<z<0.20$ for \emph{bright2}. Although they mostly refer to these non-overlapping samples, in practice for the purpose of identification of voids \citet{Sutter:2012wh} actually took each sample to include all galaxies passing the magnitude cut between redshift $z=0$ and the respective upper redshift caps, imposing the lower redshift cuts only at a later stage on the void catalogues themselves. In order to maximise our use of the data and to minimize boundary effects, we also take each sample to start form $z=0$, thus obtaining in effect four overlapping galaxy catalogues. The mean number density of galaxies in each sample is reasonably constant with redshift, though some fluctuation is present and can be corrected for. We return to this point in Sections~\ref{subsection:redshift-space} and \ref{section:ourcatalogue}.

For the LRGs we make use of the catalogue of \citet{Kazin:2010}. These authors provide two quasi-volume-limited subsamples referred to as DR7-Dim ($0.16<z<0.36$, $-23.2<M_g<-21.2$) and DR7-Bright ($0.16<z<0.44$, $-23.2<M_g<-22.8$). These two subsamples as defined by \citet{Kazin:2010} only contain LRGs in the Northern Galactic cap region of the SDSS survey; however, following \citet{Sutter:2012wh} we augment them with data from the southern Galactic sky as well.\footnote{This data is publicly available from \url{http://cosmo.nyu.edu/~eak306/SDSS-LRG.html}.} \citet{Sutter:2012wh} refer to the two subsamples thus obtained as \emph{lrgdim} and \emph{lrgbright} (note that they only apply the redshift selection $0.36<z<0.44$ to the void catalogue obtained from \emph{lrgbright}, and not to the LRGs themselves), and we adopt the same notation.

\begin{table*}
\begin{minipage}{135mm}
\caption{Details of the galaxy samples used in this work. Number densities are calculated using the comoving volumes.}
\begin{tabular}{@{}cccccc}
\hline
Sample name & Galaxy type & Magnitude limit
& Redshift extent & Number of galaxies & Mean number density  \\
& & & & & $(h^{-1}\rmn{Mpc})^{-3}$\\
\hline
\emph{dim1} & Main sample & $M_r<-18.9$
& $0<z<0.05$ & $63\,639$ & $2.4\times10^{-2}$  \\
\emph{dim2} & Main sample & $M_r<-20.4$
& $0<z<0.10$ & $178\,099$ & $8.7\times10^{-3}$  \\
\emph{bright1} & Main sample & $M_r<-21.35$
& $0<z<0.15$ & $164\,647$ & $2.5\times10^{-3}$  \\
\emph{bright2} & Main sample & $M_r<-22.05$
& $0<z<0.20$ & $77\,770$ & $5.1\times10^{-4}$  \\
\emph{lrgdim} & LRG & $M_g<-21.2$
& $0.16<z<0.36$ & $67\,567$ & $9.4\times10^{-5}$  \\
\emph{lrgbright} & LRG & $M_g<-21.8$
& $0.16<z<0.44$ & $33\,356$ & $2.6\times10^{-5}$  \\
\hline\\
\end{tabular}
\label{table:datasamples}
\end{minipage}
\end{table*}

\section{The {\small ZOBOV} algorithm}
\label{section:ZOBOV}

In this paper we will use a modified version of the parameter-free void-finding algorithm {\small ZOBOV} \citep{Neyrinck:2007gy} to identify voids in the galaxy distribution. The underlying philosophy of {\small ZOBOV} is to create a Voronoi tessellation of tracer galaxies from which the local density field can be reconstructed, and then to identify density minima. These density minima are then joined together using the watershed principle \citep*{Platen:2007qk} to create a hierarchy of voids and sub-voids.

The algorithm works as follows \citep[for a more detailed description, see ][]{Neyrinck:2007gy}. Given the coordinates of the set of points representing galaxies in any sample, we first construct a Voronoi tessellation of the survey volume. Each point $i$ is thus associated with a Voronoi cell consisting of the region of space closer to it than to any other point, and a set of neighbouring points whose positions determine the extent of $i$'s Voronoi cell, and whose cells neighbour that of $i$. We then obtain a local density estimate at the position of each galaxy based on the volume of its Voronoi cell relative to the mean volume of all cells. This is known as the Voronoi tessellation field estimator or VTFE \citep{Schaap:2007}. At this stage a redshift-dependent normalization can be applied to account for variations in the local mean number density of galaxies in the sample with redshift. 

Having obtained an estimate of the density field, {\small ZOBOV} searches for local minima of this field. Around each minimum it constructs a ``zone", consisting of all particles that -- in the terminology of the watershed transform -- form the catchment basin of the ``core" or minimum-density particle. All galaxies up to the watershed are included in the zone, so zones generally include some overdense regions at the edges. Zones are then joined together to form a hierarchy of voids as in \citet{Platen:2007qk}: each zone annexes neighbouring zones or groups of zones in ascending order of the lowest density point on the common watershed ridge separating them, but it does not annex any zone with a lower core density. Thus, barring complications such as holes in the survey region, the zone around the global minimum density particle eventually annexes all other zones to form a ``void" spanning the entire survey volume, and a complete hierarchy of sub-voids and sub-sub-voids is obtained. This may be regarded as a natural advantage, since it corresponds to the physical hierarchy that voids are thought to have in the real Universe \citep[e.g. ][]{Dubinski:1992tr,Sheth:2003py,Furlanetto:2005cc,Paranjape:2011bz}. The process of joining zones is entirely parameter-free, dependent only on the local topology of the density field, and -- in contrast to other void-finding algorithms \citep[e.g.][]{Hoyle:2001kn,Brunino:2006ym,ForeroRomero:2008ig,Foster:2009rt} -- makes no assumptions at all about the shape of voids.

At this stage, however, a parameter-dependence must be introduced as the hierarchy of voids obtained from the previous step requires severe pruning. This is primarily because {\small ZOBOV} detects very large numbers of voids, most of which are spurious fluctuations in the density field that are simply a result of shot noise \citep{Colberg:2008,Neyrinck:2007gy}. Indeed, {\small ZOBOV} treats every local minimum as a void, and thus reports many voids even in Poisson point distributions. To judge whether reported voids in the galaxy distribution are real or spurious one must therefore apply some selection criterion that is unlikely to be satisfied by voids in a Poisson distribution. In addition, the largest voids in the hierarchy can join several density minima together and become essentially arbitrarily large, while extending across intervening overdense regions. This is undesirable, so some criterion must be applied to halt the growth of voids.

One method of achieving both aims is through the use of the void \emph{density ratio}, $r$. This is defined for each void as the ratio of the minimum watershed density $\rho_\mathrm{link}$ at which it is annexed by a deeper void to the minimum density of its core particle: $r=\rho_\mathrm{link}/\rho_\mathrm{min}$. If the void is not annexed by a deeper void, $\rho_\mathrm{link}$ is taken as the lowest density particle on its boundary. \citet{Neyrinck:2007gy} estimated the fraction of voids found in Poisson distributions that have density ratio greater than $r$ to be given by
\beq
P(r)=\exp[-5.12(r-1)-0.8(r-1)^{2.8}]\,.
\eeq
It is thus possible to define a  ``statistical significance" for each void as the probability that a randomly chosen void identified in a Poisson point distribution has an equal or larger density ratio. Any subvoids which exceed a specified significance threshold can then be removed from their parent void and regarded as independent, and any voids which fail to meet the significance threshold after all additions are completed can be excluded from the final list.

Alternatively, rather than the statistical significance, some physical criteria can be chosen to select voids with desired characteristics. An advantage of such a strategy is that the information provided by the density ratio alone is limited: for instance, fewer than $1\%$ of voids in a pure Poisson distribution of points have a minimum density $\rho_\mathrm{min}<0.3\overline{\rho}$, so if such voids are found in  the data it might be inappropriate to exclude them from the final catalogue purely on the basis of a low $r$ value. On the other hand, this criterion is not well suited to deciding when to separate sub-voids from their parents. 
We therefore adopt a combination of the two approaches to create our catalogue, which is explained in detail in Section~\ref{section:ourcatalogue}. 

Note that by removing qualifying independent voids from their parents, we break the hierarchy structure and provide a set of completely disjoint, non-overlapping void regions. We regard this as an advantage, though for some purposes double-counting of void regions may be preferred \citep[e.g. ][]{Lavaux:2011yh}.

We also apply the {\small ZOBOV} algorithm to the inverse of the density field to return a complementary list of overdense structures. Such a feature is in fact already present in the {\small ZOBOV} software. In this sense it is similar to the halo-finder {\small VOBOZ} \citep*{Neyrinck:2004gj}, except that {\small VOBOZ} includes an additional step which uses particle velocity information to determine \emph{bound} particles and return virialized haloes of roughly spherical shape. Our aim is \emph{not} to produce a list of bound spherical haloes; rather we wish to find peaks in the density field analogous to the large-scale troughs returned by the void finder, and to include full volume and shape information. We describe the details of the selection procedure in Section~\ref{section:ourcatalogue}. The structures obtained are highly non-spherical and generally much larger than typical galaxy cluster dimensions. We therefore refer to them as ``superclusters" to avoid confusion. 

\section{Problems with the previous void catalogue}
\label{section:Sutter}

We now turn to the properties of the existing public catalogue of voids. A detailed description of the algorithm used to identify these voids is given in \citet{Sutter:2012wh}. Our aim in the following is to demonstrate that the data in the public catalogue is inconsistent with this stated algorithm, and with the interpretation of these structures as voids. We note that the catalogue has already undergone several revisions reflecting previous corrections and updates; we use the version dated 20 February 2013.\footnote{Available from \url{http://www.cosmicvoids.net}.} There are also two versions of the catalogue, corresponding to the use of comoving  or redshift coordinates, and we discuss differences between these two in \ref{subsection:redshift-space} below. Unless otherwise specified we use the comoving version to demonstrate inconsistencies, but our comments are equally applicable to both.

Before proceeding, we first define certain quantities that will be used below. From the galaxy catalogues described in Section~\ref{section:data}, we obtain the total number of galaxies in each strictly defined (non-overlapping) sample, $N_\mathrm{sample}$, and its total volume $V_\mathrm{sample}$. These are used to obtain the mean number density of galaxies, $\overline{\rho}$, for each galaxy sample. For all subsamples except \emph{dim1} and \emph{lrgdim}, this mean differs slightly from that calculated using the full redshift extents. From the \citet{Sutter:2012wh} public catalogue, we obtain in addition the total number of particles used in the tessellation, $N_\mathrm{total}$ (including mock particles placed at survey boundaries and in holes), and the total volume of the tessellation box, $V_\mathrm{box}$, used for each sample.\footnote{We thank Paul Sutter for explaining how to obtain $V_\mathrm{box}$.} The ratio $V_\mathrm{box}/N_\mathrm{total}$ gives the mean volume of each Voronoi cell in the tessellation.
 
We then extract the following quantities for each void in the catalogue: the total number of galaxies, $n_\mathrm{gal}$, contained within it; the total volume $V_\mathrm{void}$; the minimum density of the core particle, $\rho_\mathrm{core}$; and the void density ratio $r$. The void volume is provided both in units of $h^{-3}$Mpc$^3$ and normalized to the mean volume of a Voronoi cell in the tessellation. We refer to the latter quantity as $V_\mathrm{void}^\prime$. The core density $\rho_\mathrm{core}$ is provided in normalized units of the mean density of all $N_\rmn{total}$ Voronoi cells; using the quantities described above, this can be converted into the minimum density of each void, $\rho_\mathrm{min}$, and expressed in units of the mean number density  $\overline{\rho}$. As \citet{Sutter:2012wh} do not apply any correction for a variation in the local number density with redshift, a single value of $\overline{\rho}$ for each sample will suffice for the discussion in this section.

\begin{figure}
\includegraphics[width=85mm]{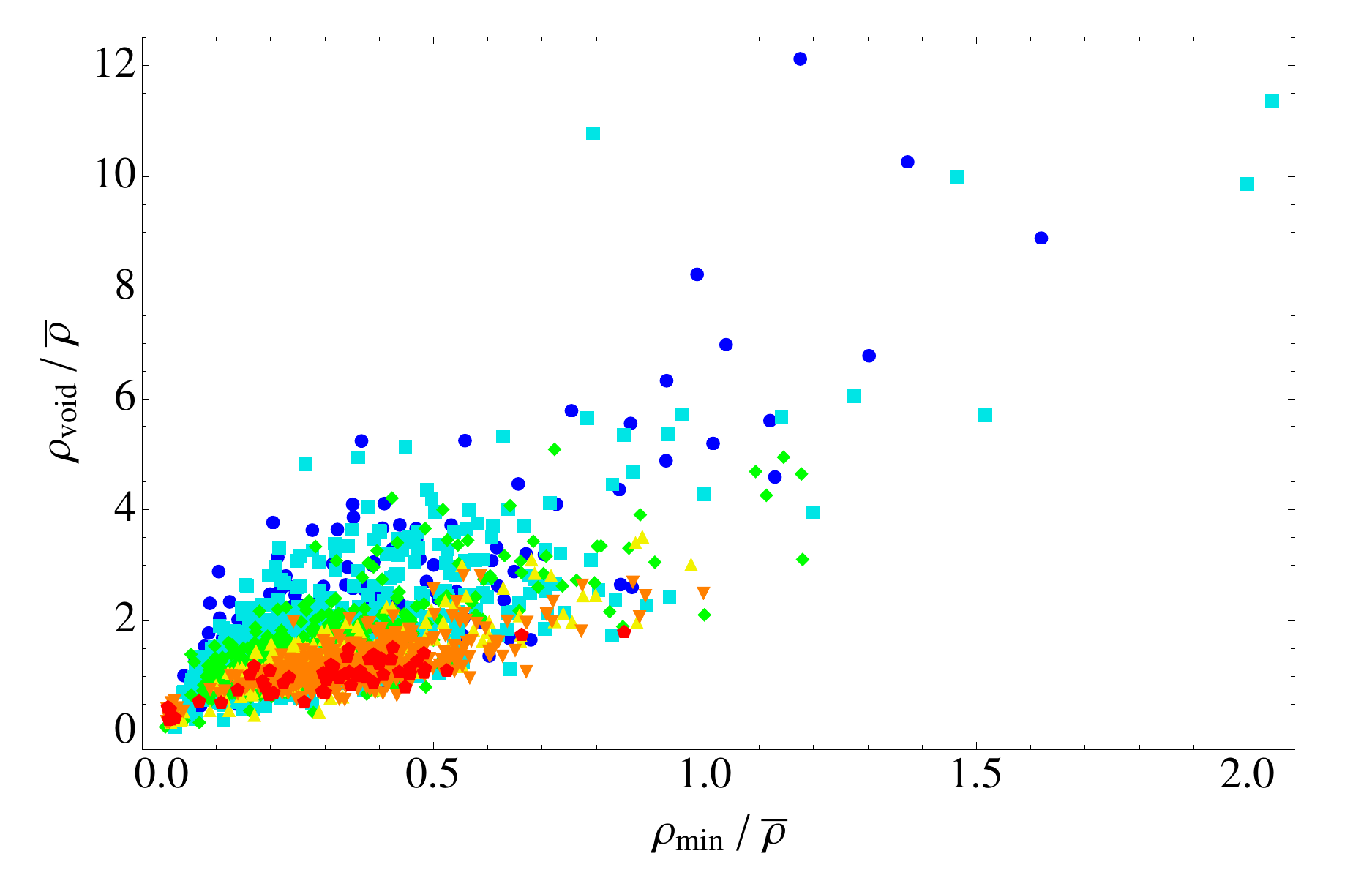}
\caption{The values of the minimum density $\rho_\mathrm{min}$ and the average density $\rho_\mathrm{void}$ of all $1985$ real-space ``voids" in the catalogue of \citet{Sutter:2012wh}. Voids identified from different samples are shown as blue circles (\emph{dim1}), cyan squares (\emph{dim2}), green diamonds (\emph{bright1}), yellow upwards triangles (\emph{bright2}), orange downwards triangles (\emph{lrgdim}) and red pentagons (\emph{lrgbright}). Both $\rho_\mathrm{min}$ and $\rho_\mathrm{void}$ are normalised to units of the mean number density for the respective samples. Most ``voids" are in fact overdense, and some dramatically so: this is inconsistent with the density cuts claimed to have been applied.} 
\label{figure:Sutterrho}
\end{figure}

\subsection{Overdense ``voids"}
\label{subsection:overdensevoids}

The first question that can be asked of the structures listed in the \citet{Sutter:2012wh} catalogue is whether they are in fact underdense with respect to the average number density of galaxies in their respective sample. To this end we obtain $\rho_\mathrm{void}=n_\mathrm{gal}/V_\mathrm{void}$ and  $\rho_\mathrm{min}$ for each void in the catalogue as described above. Since our reconstruction of $\rho_\mathrm{min}$ requires knowledge of the mean volume of each Voronoi cell in the tessellation and this quantity is not explicitly provided in the catalogue, we estimate it in two independent ways: directly, as $V_\mathrm{box}/N_\mathrm{total}$, and from the ratio $V_\mathrm{void}/V_\mathrm{void}^\prime$, which is a constant for all voids from a particular sample. We have checked that both methods give the same result. Figure~\ref{figure:Sutterrho} shows the distribution of $\rho_\mathrm{void}$ and $\rho_\mathrm{min}$ values for all voids in the catalogue.

\citeauthor{Sutter:2012wh} claim to have restricted their catalogue to include only those voids with a mean overdensity of $\delta\leq-0.8$, i.e., with $\rho_\mathrm{void}\leq0.2\overline{\rho}$. However, Figure~\ref{figure:Sutterrho} clearly shows that several of the structures in the catalogue are in fact grossly overdense, with $\rho_\mathrm{void}$ on occasion more than $10$ times the mean number density of galaxies in the sample. Only $8$ of the listed voids actually satisfy the stated mean density condition.\footnote{Following private correspondence, the authors now acknowledge on their website that the mean density criterion  stated in their paper was in fact not applied.} As discussed further in Section~\ref{subsection:voiddefinition}, the mean density of voids found by {\small ZOBOV} tends to be biased high, and so is in any case not an appropriate parameter on which to base a selection cut. However, cuts on other more appropriate parameters can help avoid the extremes of overdensity seen in the \citet{Sutter:2012wh} catalogue.

In addition, \citeauthor{Sutter:2012wh} claim to have applied a futher central density cut, to remove from their catalogue any void which contains a density contrast $\delta>-0.8$ within a specified central region. However, for  the majority of the structures listed in their catalogue, the minimum density contrast for any of the void member galaxies is $\delta_\mathrm{min}>-0.8$ (i.e. $\rho_\mathrm{min}>0.2\overline{\rho}$). Indeed for several structures the \emph{minimum} density is greater than the average.  

The reason for this second inconsistency is due to the design of the central density cut employed, which is based not on the reconstructed density field but on simply counting the number of galaxies in a central sphere about the void centre. For the vast majority of voids listed in the catalogue, this central sphere is so small that even if it were to be placed at random in a region populated at exactly the mean density, it would often contain no galaxies at all \citep{NH:2013b}. This means that the ``central density" measured by this method is not much better than Poisson noise.

\subsection{Survey boundary contamination}
\label{subsection:contamination}

\begin{figure}
\includegraphics[width=85mm]{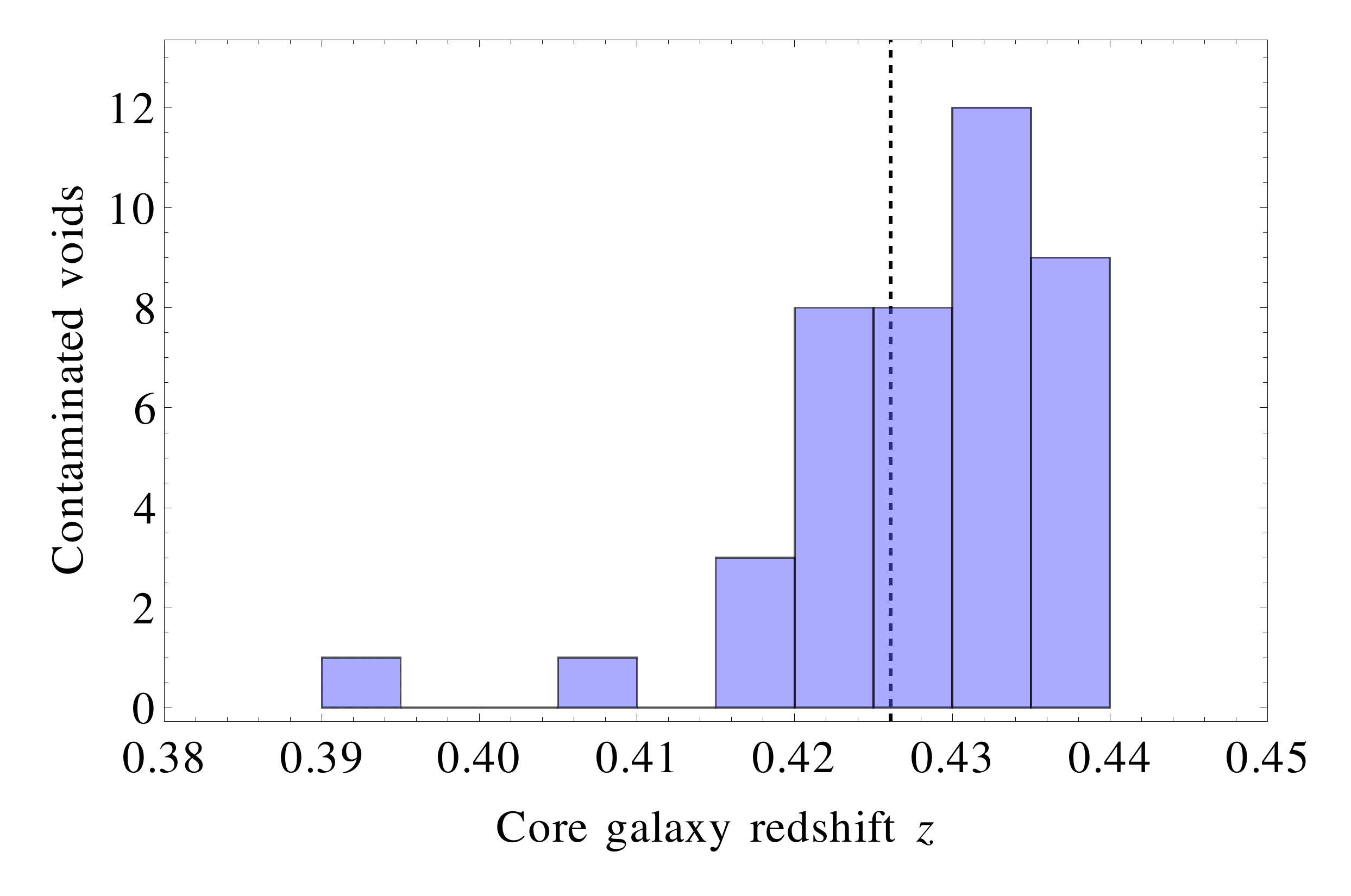}
\caption{Redshift distribution of the minimum-density or core galaxy of ``voids" identified by \citet{Sutter:2012wh} in the \emph{lrgbright} galaxy sample. Only those cores identified as being contaminated due to leakage of their Voronoi cells outside the surveyed region are shown; these account for $42$ of the $56$ listed in the catalogue. The maximum redshift extent of the \emph{lrgbright} sample is $z=0.44$. The vertical dashed line indicates the redshift corresponding to a radial distance of $\overline\rho^{-1/3}$ (i.e., approximately the mean galaxy separation) away from this maximum redshift. Contaminated cores cluster near $z=0.44$ due to the absence of containing mocks above this redshift.} 
\label{figure:contamination}
\end{figure}

The finite redshift extents of the galaxy samples used, together with survey boundaries and holes, must be accounted for during the Voronoi tessellation stage of the {\small ZOBOV} algorithm. To do this, \citet{Sutter:2012wh} place mock boundary particles along masked regions and around the survey boundary to create a thin buffer that completely encloses the survey. These particles are intended to help terminate the Voronoi cells of all galaxies near survey edges thus preventing these galaxies from being assigned arbitrarily low densities. For galaxies neighbouring boundary mocks, the volume of the Voronoi cell -- or, equivalently, the density assigned to the galaxy -- depends on the position of that mock particle and is therefore to some extent arbitrary. Such ``edge" galaxies should then be excluded from the density field determination after the tessellation stage.

However, there are two important flaws with the method \citeauthor{Sutter:2012wh} use. The first arises because they place their mock boundary particles with a fixed density of $10^{-3}\,(h^{-1}\mathrm{Mpc})^{-3}$ for all six galaxy samples. As can be seen from Table~\ref{table:datasamples}, this mock density is \emph{lower} than that of real galaxies for three of the six galaxy samples, and only significantly higher than it for the two LRG samples. For the \emph{dim1} sample in particular, the mock number density is only $\sim0.04$ times that of the galaxies. Since the boundary layer is by construction very thin, such a low volume density means that large gaps are present in the sheath of mocks enclosing the survey volume, through which Voronoi cells of galaxies can leak out of the surveyed region despite not being identified as edge galaxies. This can only be avoided by ensuring that the boundary mocks always have a significantly higher number density than that of the galaxies they enclose.

The second problem is that although their paper describes the placement of ``caps" of mock particles at both the minimum and maximum redshift of each sample, the high-redshift cap was removed in a subsequent revision of the catalogue. The absence of this cap has serious consequences for galaxies at higher redshifts, which are also assigned artificially low densities because their Voronoi cells leak out of the surveyed volume. They are not identified as edge galaxies, however, since they are not close to any mocks. The method by which \citeauthor{Sutter:2012wh} attempt to handle such cases is based on the void centre and the member galaxy positions, rather than the actual Voronoi cells, and is therefore inadequate for identifying leakage.


To estimate the extent of the survey boundary contamination in the published catalogue, we repeat the Voronoi tessellation using our own placement of boundary mocks. These are always chosen to have number density $10$ times that of the galaxies in the respective samples, and we also include a thin dense layer of mocks outside the maximum redshift extent of the sample to form a cap preventing leakage in the radial direction. As expected, the number of identified edge galaxies then increases, by up to $12\%$ of the total number of all galaxies, compared with those identified using the original mocks. Also as expected, the maximum increase is for the \emph{dim1} sample, where the original mock density was most deficient.

We then test the core or minimum-density galaxy for each of the voids listed in the \citet{Sutter:2012wh} catalogue. We find that with our more appropriate placement of mocks, an average of $~35\%$ of these density minima (over all samples) are in fact edge galaxies. This fraction does not change when considering only ``central" voids (i.e. those supposed to be far from any survey boundary). Had correct account been taken of boundary contamination effects, these galaxies would not constitute density minima, and therefore would not have formed voids.

For the lowest-redshift samples, the inadequate density of mocks makes a higher contribution to this boundary contamination than the absence of the high-redshift cap. This role reverses as the sample galaxy density decreases. Indeed for the LRG samples, a mock density of $10^{-3}\,(h^{-1}\mathrm{Mpc})^{-3}$ would normally be sufficient to prevent leakage. Instead here it is the absence of the high-redshift cap which is the major problem. Figure~\ref{figure:contamination} shows the redshift distribution of boundary-contaminated core galaxies in \emph{lrgbright}. In this sample $42$ out of the $56$ listed voids ($31$ of $44$ ``central" voids) have core galaxies whose Voronoi cells have leaked out of the survey volume. These contaminated cores can be seen to cluster at the maximum redshift extent of the sample, indicative of the problem due to the absence of boundary mocks at the high-redshift cap. 

We note that in creating a more recent catalogue of voids in the SDSS DR9 data \citep{Sutter:2013DR9}, the authors have also not used mocks above the maximum redshift extent of the sample, and therefore this catalogue will suffer from the same problem of Voronoi cell leakage at high redshifts.

\begin{figure}
\includegraphics[width=88mm]{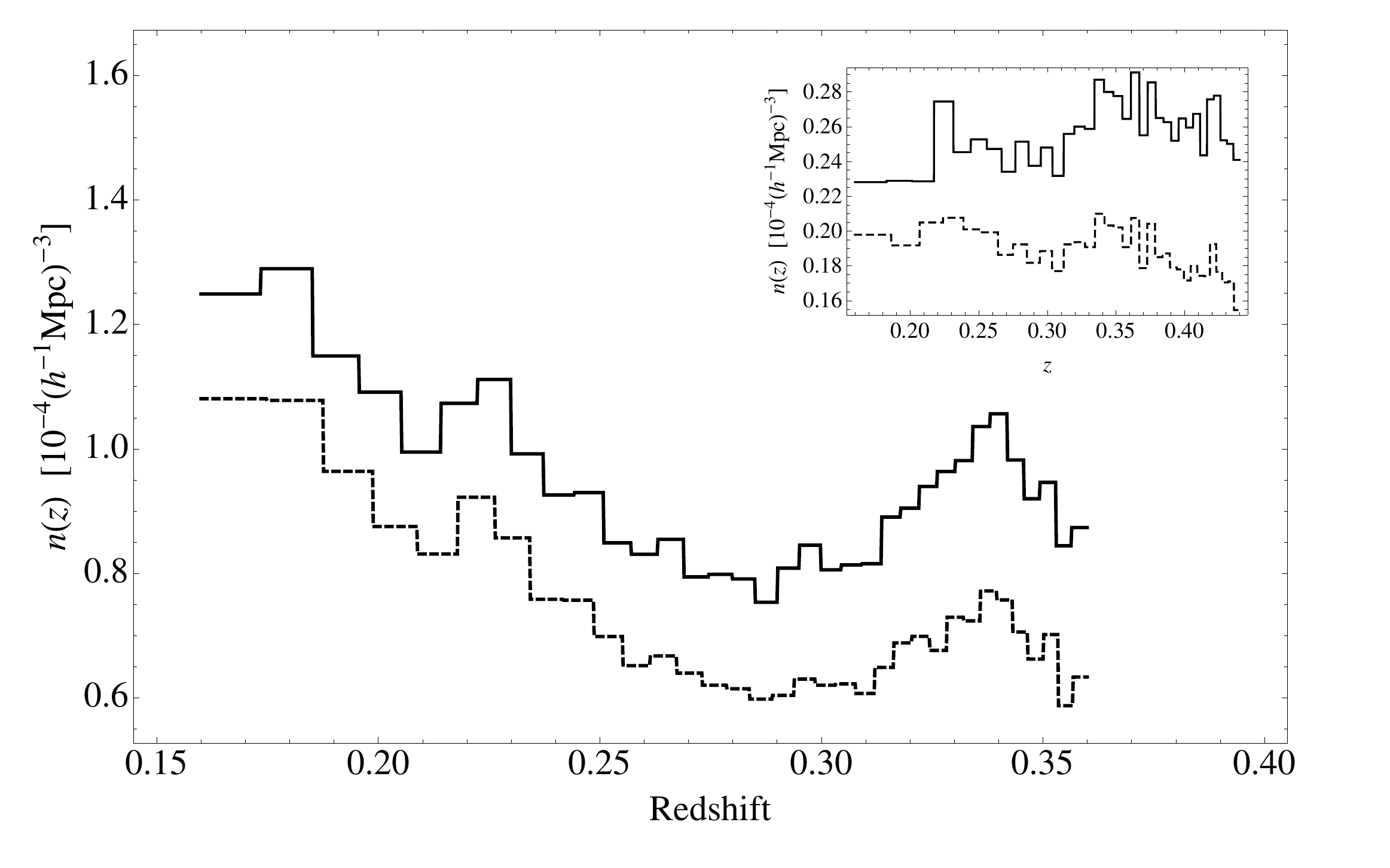}
\caption{Radial selection function for the \emph{lrgdim} sample: the number density of galaxies $n(z)$ is shown, in comoving coordinates by the solid line and in the redshift-space coordinate system defined in Equation~\ref{eq:redshiftcoords} by the dashed line. Bin widths are chosen to span equal volumes. Our algorithm is designed to correct for this radial variation in density.
\emph{Inset:} The same figure for the \emph{lrgbright} sample. Note the different behaviour in redshift coordinates.} 
\label{figure:nofz}
\end{figure}

\subsection{Statistical significance of voids}
\label{subsection:Poisson}

As discussed in Section~\ref{section:ZOBOV}, a known feature of the {\small ZOBOV} algorithm is that it reports orders of magnitude more voids than other void finders when applied to the same data \citep{Neyrinck:2007gy,Colberg:2008}, most of which are artefacts of Poisson noise. It is therefore important to check whether the characteristics of the voids listed in the \citet{Sutter:2012wh} catalogue are significantly different to those expected due to Poisson noise.

In fact the majority of voids have very low density ratios: fewer than $10\%$ of the voids have $r>2.0$, the approximate $99.3\%$ C.L. value obtained from Poisson simulations \citep{Neyrinck:2007gy}. If the stated density cuts had been correctly applied and the listed voids were extremely underdense, they might still have been regarded as real. In reality, however, $\sim40\%$ of the listed voids are neither sufficiently underdense nor have sufficiently large density ratios to be statistically distinct, at the $3\sigma$-equivalent significance level, from the spurious detections expected from Poisson noise.

\subsection{Choice of coordinate system}
\label{subsection:redshift-space}

\citeauthor{Sutter:2012wh} provide two separate void catalogues drawn from the same galaxy catalogue data, for two alternative conversions of each galaxy's sky latitude $\theta$, sky longitude $\phi$ and redshift $z$ to Cartesian coordinates. The primary catalogue uses the coordinate definitions
\bea
\label{eq:redshiftcoords}
x^\prime&=&\frac{cz}{H_0}\cos{\phi}\cos{\theta}\,,\nonumber\\
y^\prime&=&\frac{cz}{H_0}\sin{\phi}\cos{\theta}\,,\\
z^\prime&=&\frac{cz}{H_0}\sin{\theta}\,,\nonumber
\eea
where $c$ is the speed of light and $H_0$ is the Hubble parameter at redshift $z=0$, while the secondary catalogue takes the radial distance coordinate to be equal to the comoving distance $\chi(z)$ to redshift $z$ as calculated in the WMAP7 cosmology ($\Omega_M = 0.27$, $\Omega_\Lambda=0.73$ and $h=0.71$). They refer to the these two coordinate choices as ``redshift-space" and ``real-space" respectively, though in actual fact the galaxy positions are measured in redshift-space in both cases, since the radial distance is always computed as a function of observed redshift and so is subject to peculiar velocity distortions. It is therefore more appropriate to refer to ``redshift" or ``comoving" coordinate systems.

\citeauthor{Sutter:2012wh} claim that the properties of their voids are largely insensitive to the coordinate system used, but this claim is incorrect. This can be seen immediately by simply comparing the total number of voids contained in the two catalogues: while there are $1495$ voids in the primary (redshift) catalogue, there are $1985$ corresponding comoving voids, or almost a third more. For ``central" voids the corresponding numbers are $787$ and $1177$ in redshift coordinates and comoving coordinates respectively, a relative difference of almost $50\%$. Clearly these two catalogues represent statistically different populations of voids.

An immediate reason for this difference is the fact that transforming from comoving coordinates to those in Equations~\ref{eq:redshiftcoords} changes the redshift-dependence of the mean number density of galaxies (see Figure~\ref{figure:nofz}). The criteria for selecting voids are sensitive to this mean, but  \citeauthor{Sutter:2012wh} do not account for its variation. 

Note that even if the two populations of voids were statistically the same, it is not at all clear that there is any simple correspondence between individual voids identified in the two coordinate systems, and properties such as the radius and volume of voids will certainly change. It is therefore puzzling that some previous studies looking at effects dependent on the comoving size of voids  \citep{Ilic:2013cn, Planck:ISW,Melchior:2013} have used the version of the catalogue in redshift coordinates.

We return to this issue in Section~\ref{section:results} once the other problems with the \cite{Sutter:2012wh} methodology identified above have been corrected.



\section{Construction of a self-consistent superstructure catalogue}
\label{section:ourcatalogue}

The various problems with the existing catalogue documented above mean that an entirely new and self-consistent catalogue is required, which we construct as described below. In addition we construct a catalogue of superclusters obtained by applying an inversion of the {\small ZOBOV} algorithm (i.e., applying {\small ZOBOV} to the inverse of the reconstructed density field). The catalogues are constructed using the SDSS main galaxy and LRG samples described in Section~\ref{section:data}. 

We provide two versions of the catalogue. For the primary one we convert the sky positions and redshifts of galaxies into Cartesian coordinates in comoving space as: $x^\prime = \chi(z)\cos(\theta)\cos(\phi),\;y^\prime = \chi(z)\cos(\theta)\sin(\phi),\;z^\prime=\chi(z)\sin(\theta)$, where $\chi(z)$ is the comoving distance to redshift $z$, calculated using a WMAP7 cosmology ($\Omega_M = 0.27$, $\Omega_\Lambda=0.73$ and $h=0.71$). We also make available for download a secondary catalogue using the same redshift coordinates as \citet{Sutter:2012wh}; however, we attach strong caveats to the use of the redshift coordinate catalogue, as discussed in Section~\ref{subsection:redshiftstructures}.

\begin{figure*}
\includegraphics[width=120mm]{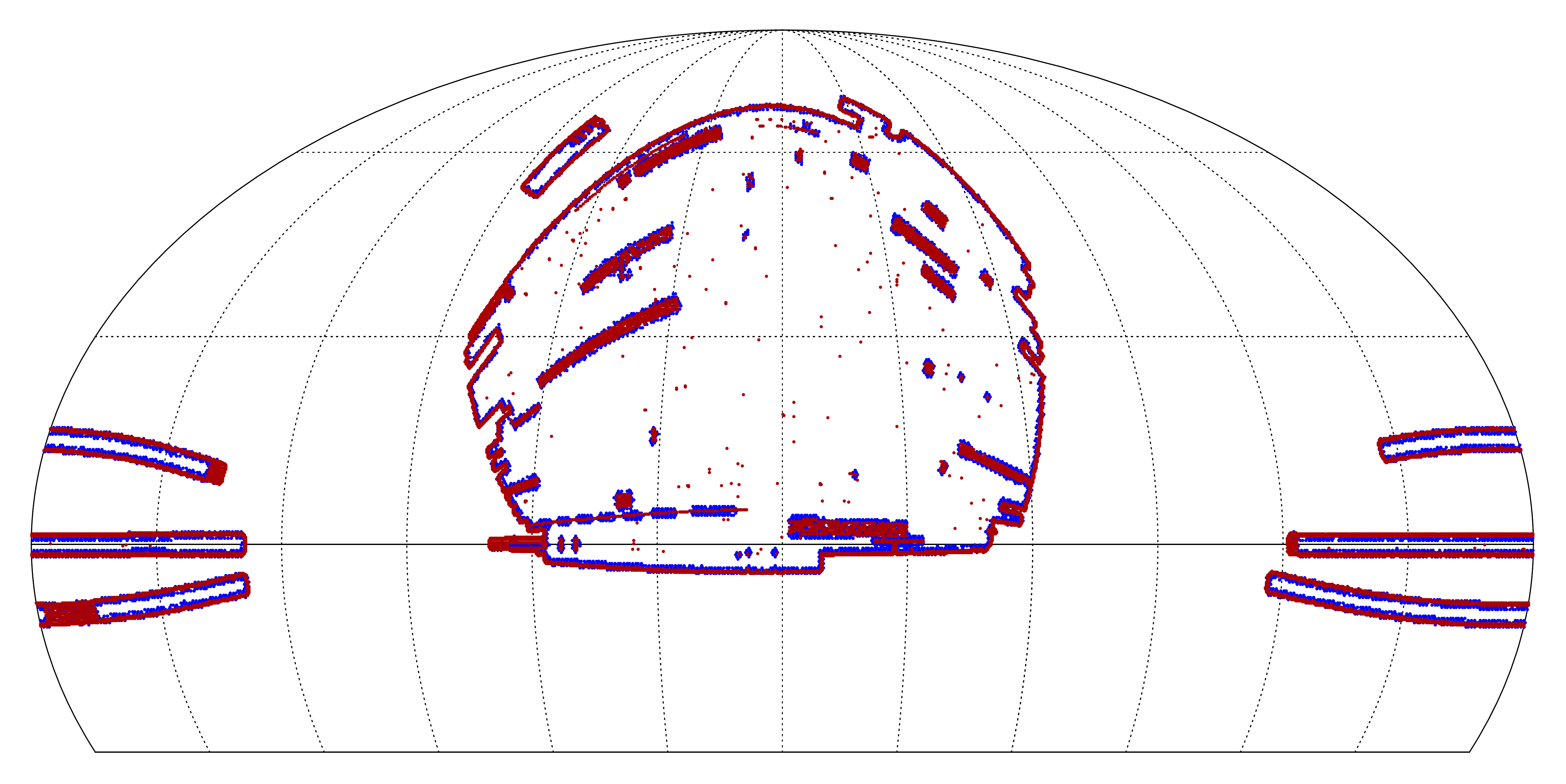}
\caption{Mollweide projection of the SDSS survey boundary and bright star mask in a {\small HEALPix} $N_\rmn{side}=512$ pixelization. The red points show the locations of boundary pixels neighbouring but outside the surveyed area, as determined from the \emph{safe0} mask. These are used for the placement of mock boundary particles. The blue points beneath them show the boundary pixels used by \citet{Sutter:2012wh}. The boundaries do not exactly overlap as they did not use the same mask, though the reasons for this are unclear.} 
\label{figure:surveymask}
\end{figure*}

\subsection{Reconstructing the density field}
\label{subsection:density}

The first step in identifying superstructures is to build a Voronoi tessellation of each galaxy catalogue from which the local density field can be reconstructed. To do this correctly, one must account for the finite volume of the catalogue due to the survey boundaries and redshift cuts, as well as the presence of ``holes" due to masked regions within the survey volume. We therefore start with a {\small HEALPix} \citep{Gorski:2004by} pixelization of the survey window and mask at $N_\rmn{side}=512$, obtained by conversion of the DR7 \emph{safe0} mask \citep{Blanton:2004aa} to {\small HEALPix} format using the {\small MANGLE} software \citep{Swanson:2008}.\footnote{This data can be downloaded from the {\small MANGLE} website at \url{http://space.mit.edu/~molly/mangle/download/data.html}.} Although this procedure is ostensibly the same as that used by \cite{Sutter:2012wh}, we obtain a somewhat different pixelization of the survey window. In particular, the mask used by \citeauthor{Sutter:2012wh} appears to have a different resolution and does not account for all the holes in the survey region (see Figure~\ref{figure:surveymask}).

We then identify pixels near the survey boundaries and within these pixels we place mock boundary particles, distributed randomly in the volume along the redshift extent of the survey, at $10$ times the sample mean number density for each sample. In addition, we place a similarly dense layer of mock particles just outside the maximum and, where applicable, the minimum redshift extents of each sample in order to create redshift caps. These boundary particles taken together create a thin layer of particles completely enclosing the survey galaxies, which ensures that the Voronoi cells of galaxies near the survey edges are not allowed to extend arbitrarily far outside the survey volume.

Note that our method of handling boundary particles differs from that of \cite{Sutter:2012wh}, who use a fixed boundary particle number density of $10^{-3}\;(h^{-1}\rmn{Mpc})^3$ for all samples and do not provide a layer of mocks at the maximum redshift extent. As discussed in the previous section, this provides inadequate protection against leakage of Voronoi cells outside the survey volume.

Finally, we define a cubic box which contains all the galaxies and boundary particles, which is used as an input for the tessellation step of {\small ZOBOV}. Since in some directions the box boundary is very far from the nearest galaxy, for reasons of stability of the code we needed to add additional mock particles in a sparse regular grid to fill the remaining space within the box before performing the tessellation.

After the tessellation stage we examine the adjacencies of all the real galaxies and classify any galaxy with one or more mock neighbours as an ``edge" galaxy. We find a much higher number of edge galaxies than \citet{Sutter:2012wh}, amounting to $\sim40$--$50\%$ of the total number, depending on the sample. We then remove all mock particles, their associated Voronoi cells, and information of their adjacencies. The volumes of the remaining Voronoi cells are normalised in units of the mean Voronoi volume for all real galaxies and stored for later use.

We then apply a different procedure for calculating the density at each galaxy depending on whether it is classified as an edge galaxy or not. All edge galaxies have their density set to infinity in order to prevent them from forming the core particle of any zones. The remaining non-edge galaxies are assigned a density based on inverse of the Voronoi volume (since smaller cells correspond to more densely packed galaxies). This defines the density field which is used in the zoning stage. {\small ZOBOV} uses the density normalized in units of the mean number density of galaxies, and  we include a correction for the redshift-dependence of the mean measured in redshift bins. This correction is particularly important for \emph{lrgdim}, where the mean number density varies significantly with redshift \citep{Kazin:2010}, and in redshift coordinates (see Figure~\ref{figure:nofz}), though it is also required for other samples. 

For supercluster identification, rather than modifying the {\small ZOBOV} algorithm we simply apply it to the inverse of the density field. That is, edge galaxies are assigned zero density and for remaining non-edge galaxies we use the normalised volume of the Voronoi cell (with redshift-dependent correction applied).

\subsection{Creating zones}
\label{subsection:zones}

The procedure outlined above for reconstructing the density field given a complex survey geometry is robust. In addition, as it accounts for variable selection functions, it can be applied with minimal modification to any survey, even if it is not (quasi-) volume-limited. However, two further complications must be dealt with during the process of creating zones.

Firstly, although edge galaxies have been assigned infinite densities and so cannot form the core particle around which a zone grows, the watershed algorithm naturally assigns every edge galaxy as a member of \emph{some} zone. We flag such zones as ``edge zones" and classify any void containing one or more edge zones as an ``edge void". However, we do not count the edge galaxies amongst the members of such zones and voids, nor include the contributions of their Voronoi cells to the total volume of the void. When using such edge voids, the user should be aware that their true extents have been somewhat truncated by the presence of the survey boundary.

Secondly, the complex survey geometry means that occasionally an isolated zone or void does not link to any other neighbouring one except via edge galaxies, meaning that $\rho_\mathrm{link}$ for such a zone becomes infinite. Luckily such cases are rare and we were able to remove almost all of them by requiring that to be considered any seed zone must be composed of at least $5$ galaxies. The same procedure is followed with the inverse of the density field in the case of supercluster identification, except that here it was found necessary to require that seed zones contain a minimum of $10$ galaxies.

\begin{table*}
\begin{minipage}{133mm}
\caption{Numbers of structures identified in different galaxy samples. The numbers in parentheses refer to structures found in the redshift coordinates of Equation~\ref{eq:redshiftcoords}.}
\begin{tabular}{@{}lrrrrrrr}
\hline
Structure & \emph{dim1} & \emph{dim2}
& \emph{bright1}
& \emph{bright2} & \emph{lrgdim} & \emph{lrgbright} & Total \\
\hline
Basic Type voids & 262 (238) & 676 (652) & 712 (696) & 398 (373) & 349 (376) & 193 (166) & 2590 (2501) \\
Type1 voids& 80 (80) & 271 (284) & 262 (256) & 112 (99) & 70 (63) & 13 (13) & 808 (795)\\
Type2 voids & 53 (51) & 199 (200) & 163 (169) & 70 (69) & 19 (26) & 1 (1) & 505 (516)\\
Superclusters & 419 (412) & 1192 (1176) & 896 (895) & 325 (330) & 196 (189) & 39 (43) & 3067 (3045)\\
\hline\\
\end{tabular}
\label{table:voidnums}
\end{minipage}
\end{table*}

\subsection{Defining basic voids}
\label{subsection:voiddefinition}

After the creation of the full void hierarchy by the {\small ZOBOV} watershed algorithm, it is necessary to post-process the output to obtain a usable list of voids, as discussed above. For this purpose, we must first define what constitutes a ``void".

A common definition, based on the expectations of the critical density threshold from a simple spherical evolution model for voids \citep{Suto:1984,Fillmore:1984wk,Bertschinger:1985nj}, and employed in excursion set models \citep[e.g.][]{Sheth:2003py, Paranjape:2011bz,Jennings:2013nsa}, is that a ``void" consists of a region of space with mean density $\rho_\mathrm{void}\leq0.2\overline{\rho}$. However, this criterion applies to the \emph{total} matter density field. When dealing with the reconstructed density field of different sets of biased tracers, as we do here, it is no longer clear what an appropriate value of $\rho_\mathrm{void}$ would be. 

\citet{Sutter:2012wh} chose to use a similar definition, specifying among other things that voids in the galaxy catalogues have mean overdensity $\delta_\mathrm{void}\leq-0.8$. However, as discussed in Section~\ref{section:Sutter}, this condition was not implemented: according to the information provided in the catalogue, no more than $8$ of the ``voids" they report actually satisfy this condition. In fact, after applying our improved algorithm that correctly removes survey boundary contamination and accounts for the variation of the selection function, we find that \emph{none} of the density minima found by {\small ZOBOV} in \emph{any} of the galaxy samples satisfy such a stringent condition. 

Indeed, this is not surprising: the very nature of the watershed algorithm means that the galaxies in an overdense filament or wall separating two voids are necessarily allocated to one of the two voids, or more commonly split between both. This rasies the average density of each void, such that $\rho_\mathrm{void}/\overline{\rho}$ is always $\sim1$ and occasionally rather larger, even if its centre does indeed contain a large, extremely underdense region \citep[this point is also made by][]{Achitouv:2013}. A selection cut on $\rho_\mathrm{void}$ is therefore in principle misconceived. If a strict upper limit on $\rho_\mathrm{void}$ is desired, then the use of a watershed voidfinder such as {\small ZOBOV} is probably not optimal. On the other hand, given the various other advantages {\small ZOBOV} provides, we prefer to apply different criteria to obtain a list of voids.

Choosing a particular set of criteria to halt the growth of the void hierarchy and extract a usable catalogue naturally introduces a degree of subjectivity to the final result. We therefore choose a very broad definition in the first instance. The minimal criteria we employ are:
\begin{enumerate}
\item to qualify as the starting seed for any void, a zone must have $\rho_\mathrm{min}<1$ (in units of the mean)
\item a zone or group of zones cannot be added to a void if the minimum linking density between them is $\rho_\mathrm{link}>1$, and
\item any zone or group of zones which has density ratio $r>2$ (a $3\sigma$ criterion) is not added to a deeper neighbour.
\end{enumerate}
The growth of each qualifying seed zone proceeds by addition of neighbouring zones in increasing order of their $\rho_\rmn{link}$ values until either condition (ii) or (iii) is violated. We refer to structures built according to these three conditions as ``Basic Type" void candidates. The first condition ensures that the density minimum about which the starting zone is constructed is underdense; this removes any spurious local minima in highly overdense regions that arise purely due to noise. The second condition ensures only that voids do not grow across overdense ridges. Note that voids may still contain some overdense galaxies at the edges -- indeed, the very nature of the watershed algorithm for the creation of zones means that most will do -- but they cannot expand past such an edge. The third condition stipulates that any sub-void that has less than a $0.3\%$ chance of having occurred in a Poisson distribution judged on the density ratio alone \citep{Neyrinck:2007gy} is regarded as independent of its parent void. Note that all these criteria are applied at the post-processing stage, and we do not modify the main watershed algorithm in {\small ZOBOV}.

These criteria are by construction very lenient, and Basic Type voids indeed include many with $\{\rho_\mathrm{min},\,r\}$ values that are entirely consistent with those found for voids in Poisson distributions (see Figure~\ref{figure:rhominandr}). We emphasize once again that our aim here is to provide as comprehensive a list of candidates as possible, from which users may extract their own strictly defined subsets according to the particular purpose required, but the Basic Type catalogue itself should not be used without some modification. Two possible stricter definitions are described below.

\subsection{Type1 and Type2 voids}
\label{subsection:TypeNv}

The simplest way to modify the list of Basic Type void candidates would be to impose a strict Poisson significance cut on the density ratio, e.g. $r\geq2$, to extract a subset of voids. This corresponds to the procedure applied in the void catalogue produced by \citet{Granett:2008ju}. However, we prefer not to do this for two reasons. Firstly, we find a number of voids with $\rho_\mathrm{min}<0.3\overline{\rho}$ -- which is unlikely at $>99.5\%$ C.L. for voids in Poisson simulations -- that have $r<2$, so would be unnecessarily excluded. In addition, we also find a number of voids that have relatively large values of $\rho_\mathrm{min}\sim0.6\overline{\rho}$ but also large density ratios, $r>2$. Such structures generally contain too many overdense galaxies to correspond to an intuitive understanding of what a void should be, and indeed often have the largest average densities, $\rho_\mathrm{void}\gg\overline{\rho}$. 

We therefore define ``Type1" voids as being the subset of Basic Type voids that have $\rho_\mathrm{min}<0.3\overline{\rho}$, irrespective of their density ratios. This amounts to a modification of condition (i) only.

It is also possible to modify the Basic Type catalogue such that voids are not allowed to grow across any links which are not substantially underdense. We define ``Type2" voids by modifying both conditions (i) and (ii) so that seed zones must have $\rho_\mathrm{min}<0.2$ to qualify as voids, and that zones cannot be added to an existing void if $\rho_\mathrm{link}>0.2$. Type2 voids thus match as closely as possible the usual void definition discussed in Section~\ref{subsection:voiddefinition}, since every reported void contains a contiguous central region with an overdensity $\delta\leq-0.8$ although its overall density is higher. If desired, Type2 voids could be ``trimmed" by removing member galaxies from their overdense boundaries until the condition intended by \citet{Sutter:2012wh} is satisfied. However, we do not attempt such a strategy here. 

We emphasise again that contrary to the situation with the existing catalogue, all our Type1 and Type2 voids are statistically significantly different from those expected to arise from Poisson noise.

For each identified void, we define the centre to be the volume-weighted barycentre of its member galaxies,
\beq
\mathbf{X}_v = \frac{1}{\sum_iV_i}\sum_i\mathbf{x}_iV_i\,,
\eeq
where $V_i$ is the Voronoi volume of the $i^\mathrm{th}$ member galaxy with position vector $\mathbf{x}_i$. In addition, we provide the total volume of the void, $V$, defined as the sum of Voronoi volumes of member galaxies, and its effective radius, defined as the radius of a sphere that has the same volume as that of the void:
\beq
R_\mathrm{eff} = \left(\frac{3}{4\upi}V\right)^{1/3}\,.
\eeq

\subsection{Defining superclusters}
\label{subsection:clusterdefinition}

Our philosophy in defining superclusters is similar to that employed in choosing Type1 voids. In order to break the cluster hierarchy, we employ a threshold on the density ratio $r$ of subclusters (now defined as the ratio $\rho_\rmn{max}/\rho_\rmn{link}$) to decide whether to include them as part of their parent clusters or treat them as independent. We then impose a cut on the maximum density $\rho_\rmn{max}$ in order to exclude from the final catalogue those overdense structures that have a significant likelihood of occurring in a Poisson distribution of points. However, simple volume considerations mean that smaller overdense structures are far more numerous than voids, both in Poisson simulations and in real data. We are therefore able to choose stricter significance cuts on both $r$ and $\rho_\rmn{max}$, corresponding to the $4\sigma$-equivalent confidence level, based on our own simulation results and those of \cite{Neyrinck:2004gj}. In summary, the criteria employed for selection of superclusters are:
\begin{enumerate}
\item to qualify as the starting seed for any supercluster, a zone must have $\rho_\rmn{max}>22$ (in units of the mean)
\item a zone or group of zones cannot be added to a supercluster if the maximum linking density between them is $\rho_\rmn{link}<1$, and
\item any zone or group of zones which has density ratio $r>16.3$ is not added to a denser neighbour.
\end{enumerate}

\begin{figure}
\includegraphics[width=85mm]{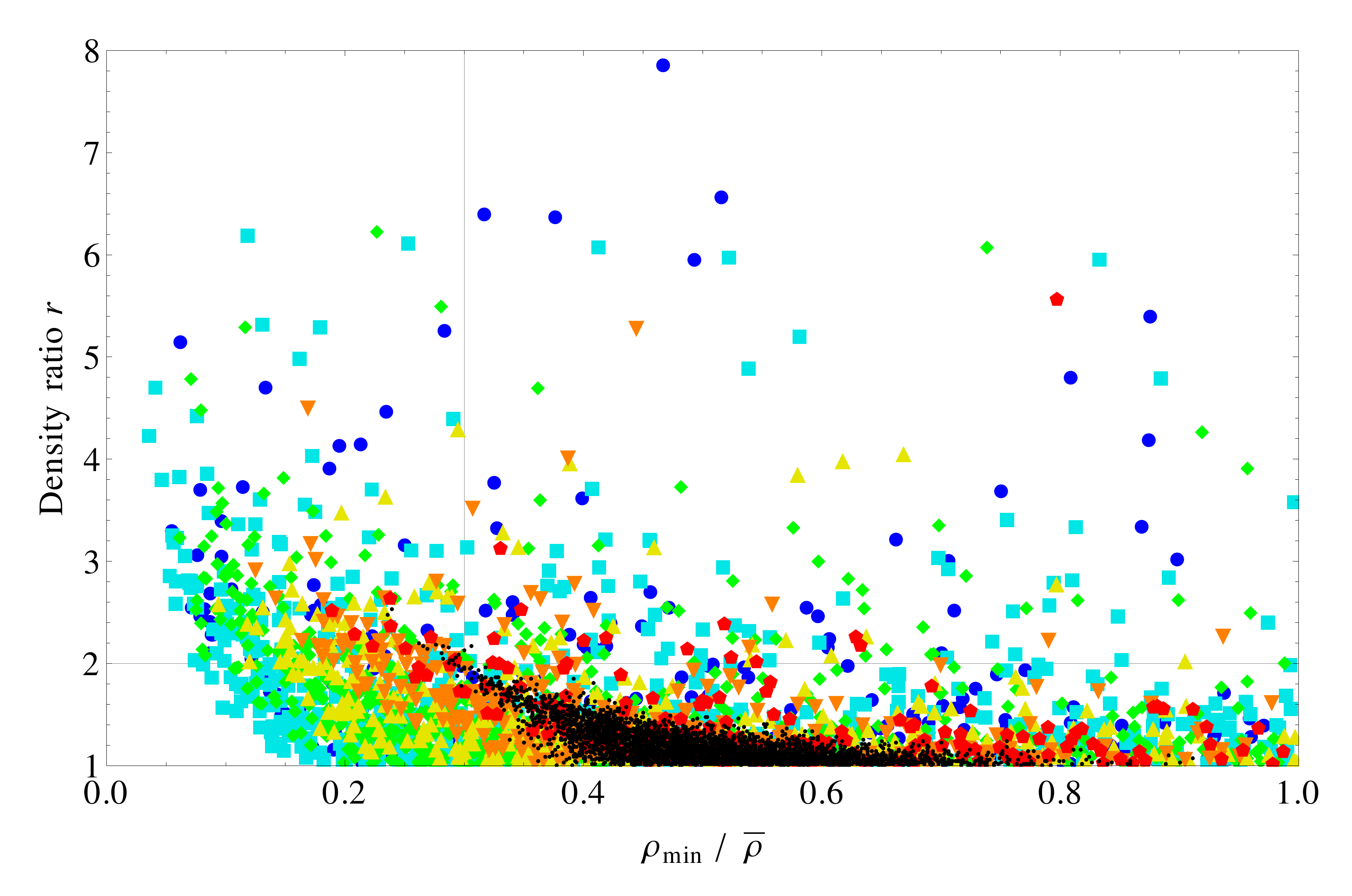}
\caption{Distribution of $\rho_\mathrm{min}$ and $r$ values for Basic Type voids (see text) found in the SDSS galaxy samples. Different samples are denoted by the same symbols as in Figure~\ref{figure:Sutterrho}. The overlaid small black points represent values for voids found in a Poisson distribution of $128^3$ points in a box, though for clarity only a tenth of these values, chosen at random, are shown. The $3\sigma$ Poisson-significance cut suggested by \citet{Neyrinck:2007gy} would keep only voids above the horizontal line; in our Type1 sample we choose instead to keep all voids to the left of the vertical line.} 
\label{figure:rhominandr}
\end{figure}

\section{Results}
\label{section:results}

\subsection{Superstructure statistics}
\label{subsection:statistics}

Table~\ref{table:voidnums} shows the numbers of voids and superclusters found in the different galaxy samples. The number of void candidates in the Basic Type catalogue is sharply reduced by the strict $\rho_\rmn{min}$ cuts applied for Type1 and Type2 voids. The decrease in numbers from Type1 to Type2 is not so severe in part because the condition on $\rho_\rmn{link}$ increases the splitting of underdense regions into separate voids. It is also clear that the reduction in numbers caused by the $\rho_\rmn{min}$ cuts is greater for the higher redshift samples. This is because the distribution of zone minimum densities shifts to higher values for these samples, as can be seen also in Figure~\ref{figure:rhominandr}. This accords with the intuitive understanding that voids evolve to become emptier with time. However, we caution that it may also be partly due to the fact that the number density of the galaxy tracers decreases by three orders of magnitude from \emph{dim1} to \emph{lrgbright}. The more highly biased LRGs do not trace smaller-scale fluctuations in the matter density so well, meaning that in some cases several unresolved smaller voids may be reported together as a single larger void. Table~\ref{table:voidvolfrac} summarises the fraction of the sample survey volume contained within comoving-space voids of different types. Note the sharp drop-off in volume fractions at higher redshifts with the more restrictive Type2 definition.

A revealing comparison can be made between the numbers of voids in Table~\ref{table:voidnums} and those obtained by \citet{Sutter:2012wh}. Despite the fact that our criteria for defining Type1 and Type2 voids are both in theory more lenient than theirs -- especially so for Type1 -- we find far fewer voids. In the case of Type2 there are a quarter as many comoving-space voids in our catalogue, and even this comparison is slightly misleading, since we keep all voids found in the full redshift extent of all the samples whereas \citeauthor{Sutter:2012wh} divide the main galaxy and LRG samples into non-overlapping bins, thereby discarding many voids.

\begin{table}
\begin{centering}
\caption{Survey volumes and void volume fractions}
\begin{tabular}{@{}lrccc}
\hline
Name & Sample volume & \multicolumn{3}{c}{Void volume fraction} \\
 & $(10^6\,h^{-3}\rmn{Mpc}^3)$ & BasicType & Type1 & Type2 \\
\hline
 \emph{dim1} & 2.7 & 0.30 & 0.26 & 0.22 \\
\emph{dim2} & 20.5 & 0.40 & 0.36 & 0.31 \\
\emph{bright1} & 67.1 & 0.42 & 0.37 & 0.26 \\
\emph{bright2} & 153.7 & 0.40 & 0.33 & 0.16 \\
\emph{lrgdim} & 720.1 & 0.42 & 0.32 & 0.07 \\
\emph{lrgbright} & 1298.6 & 0.38 & 0.23 & 0.004\\
\hline
\end{tabular}
\label{table:voidvolfrac}
\end{centering}
\end{table}

\begin{figure*}
\includegraphics[width=120mm]{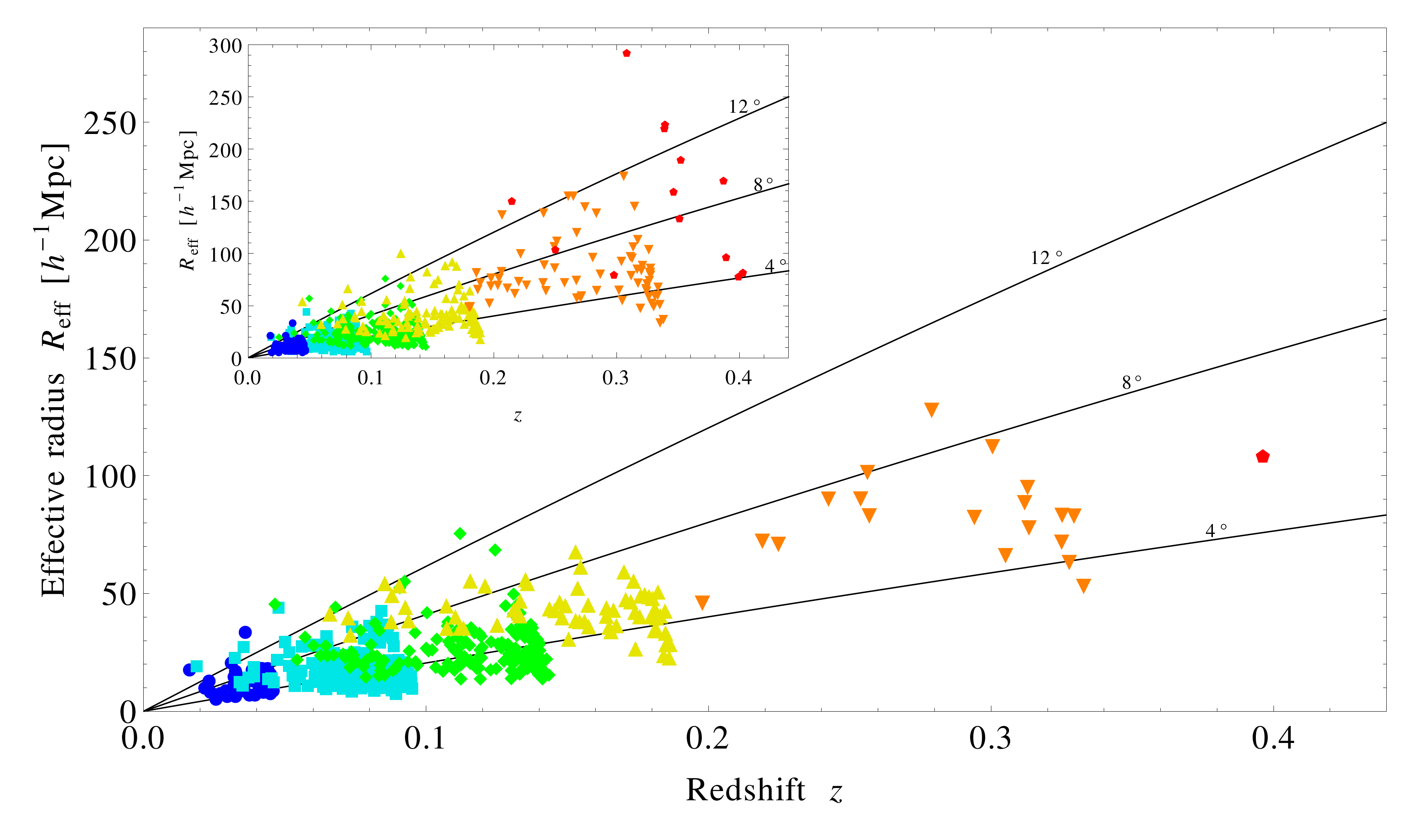}
\caption{Effective radii of Type2 voids as a function of their redshifts. Colours and symbols are as in Figure~\ref{figure:Sutterrho}. The black curves indicate angular sizes on the sky for comparison. Average void sizes increase with the mean redshift of the galaxy sample.
\emph{Inset:} The same figure for Type1 voids. These are more numerous and typically span a greater range of sizes.} 
\label{figure:radiusredshift}
\end{figure*}

\subsection{Structures in redshift coordinates}
\label{subsection:redshiftstructures}

Having corrected the shortcomings of the previous void-finding methodology, we are now able to reassess the role played by the change in coordinate system. Table~\ref{table:voidnums} shows the numbers of structures found in both coordinates. Overall, the population sizes are much closer than those found by \cite{Sutter:2012wh}. 

However, it is still not completely clear whether there is a simple relationship between structures found in the two coordinate systems. For a start, the numbers for individual samples fluctuate up and down, and the relative differences depend also on the void type definition. This is probably due to small changes introduced to individual galaxy Voronoi cells by the coordinate transformation, which affects the numbers passing the cuts on $\rho_\rmn{min}$. 

Even when the gross numbers are the same (e.g., for \emph{dim1} Type1 voids) we find that fewer than half the zone minima correspond to the same particles or locations, and even in these cases, additional properties such as the number of member galaxies of the core zone and the number of zones merged to form a void are not the same. As a result, properties such as the extent of the void and the location of its  barycentre also change. Consequently, voids found in redshift coordinates cannot in general be easily identified with underdensities in the comoving galaxy number density, nor superclusters with overdensities. It is then not clear what physical meaning to attach to structures found in these redshift coordinates.

We believe this shows that the topology of the density field -- by which we mean the location of its critical points, which is reflected in the location and extent of the zones reconstructed by the Voronoi tessellation -- is \emph{not} preserved, and so contrary to previous assertions \citep{Lavaux:2011yh,Sutter:2012wh}, the watershed void finding algorithm is not robust to this coordinate transformation. We therefore do not discuss the results in redshift coordinates any further in this paper. For the sake of completeness, we make the catalogue in redshift coordinates available for download, but we suggest that users exercise caution when using it.

\subsection{Superstructure sizes}
\label{subsection:sizes}

Figure~\ref{figure:radiusredshift} shows the distribution of effective radii of Type1 and Type2 voids identified in the different galaxy samples, as a function of the redshift of the void barycentre. For both types, the typical radius increases and large voids become more common as we move to higher redshifts. The median effective radii for Type1 voids in each sample are approximately $9\;h^{-1}$Mpc, $13\;h^{-1}$Mpc, $21\;h^{-1}$Mpc, $37\;h^{-1}$Mpc, $73\;h^{-1}$Mpc and $150\;h^{-1}$Mpc in order of increasing redshift of the samples. The corresponding maximum effective radii are $33.1\;h^{-1}$Mpc, $43.7\;h^{-1}$Mpc, $75.1\;h^{-1}$Mpc, $99.4\;h^{-1}$Mpc, $173.2\;h^{-1}$Mpc and $291.3\;h^{-1}$Mpc, indicating highly skewed distributions.

Naively, one might seek to explain this as a volume effect: if larger voids are less common per unit volume than smaller voids, then one will naturally only find large voids in the samples covering the largest volumes. However, if this were the only factor, the higher redshift samples would contain a preponderance of small voids as well as the occasional giant void, which is clearly not the case. Instead a more important factor appears to be the mean density of the galaxy tracers, which affects the minimum void size resolution achievable with {\small ZOBOV}. Thus in the sparser samples several nearby small voids may be seen as a single large void. This is also borne out by the observation that in the regions where two different galaxy samples overlap, the voids found in the sparser galaxy distribution are larger and less numerous. We find that the minimum void radius in a sample increases in roughly constant proportion to the mean inter-galaxy separation of the sample as estimated by $\overline{\rho}^{-1/3}$. (In the Basic Type catalogue, occasional examples of voids with $R_\rmn{eff}<\overline{\rho}^{-1/3}$ are found, but these are naturally removed by the selection cuts designed to eliminate spurious voids due to Poisson noise.)

It should be noted that a very large majority of the voids presented in our catalogue are ``edge" voids, meaning that one or more of their member galaxies are edge galaxies adjacent to a boundary particle in the Voronoi tessellation. The same is true for about two-thirds of the superclusters. The reason for this is the high percentage of galaxies in each sample that are edge galaxies, which is a consequence of the highly complex boundary and the presence of many bright star holes in the mask. Our treatment of edge particles means that the density field away from the edges and the location of its extrema have been conservatively reconstructed and are therefore free of boundary contamination. However, as we do not include edge galaxies in structures, it is likely that in at least some cases edge voids and clusters have been truncated by the survey boundary and have true extents larger than those reported here.

\begin{figure}
\includegraphics[width=88mm]{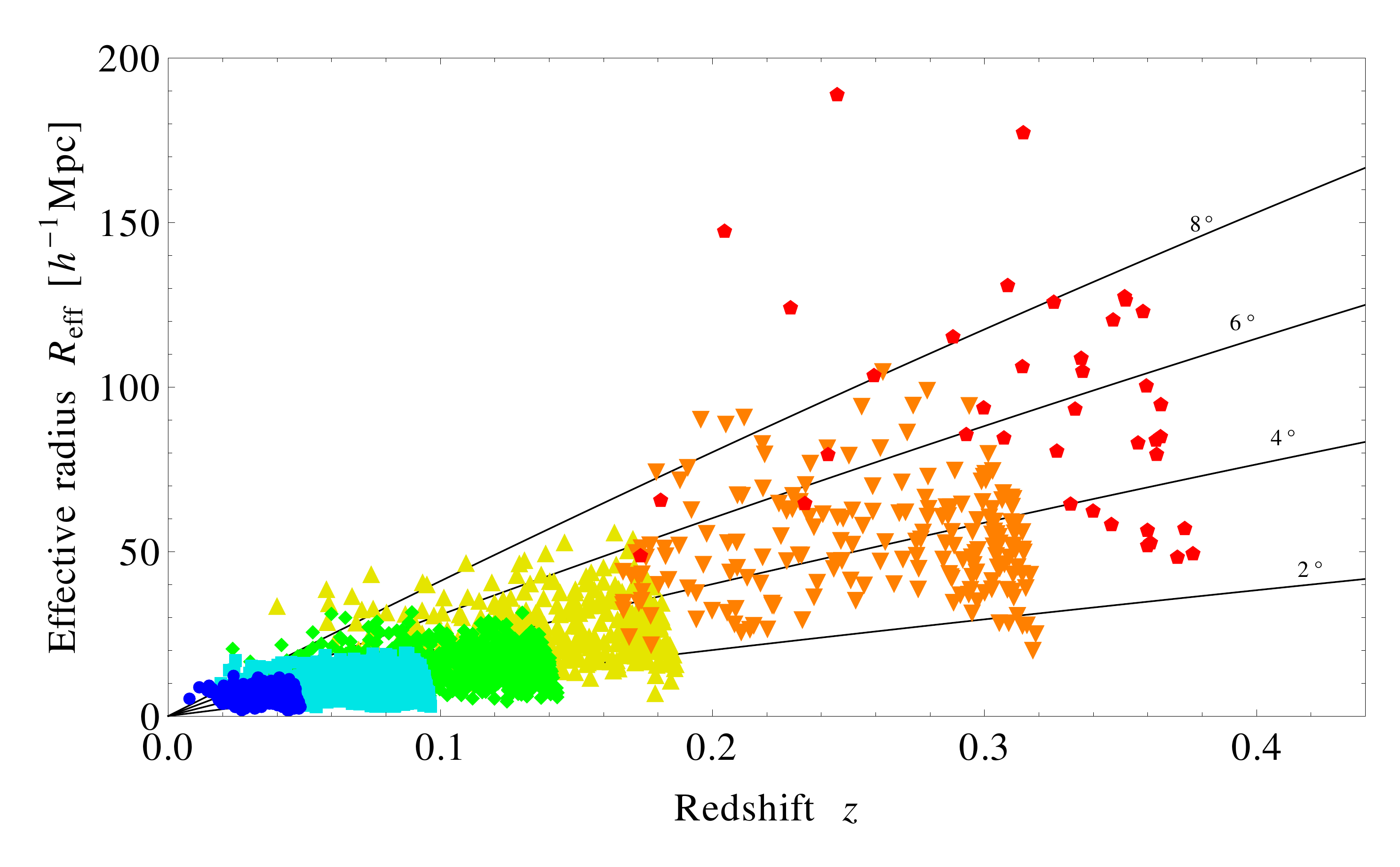}
\caption{Effective radii of superclusters as a function of their redshifts. Colours and symbols are as in Figure~\ref{figure:Sutterrho}. } 
\label{figure:clusradiusredshift}
\end{figure}

\begin{figure*}
\includegraphics[width=115mm]{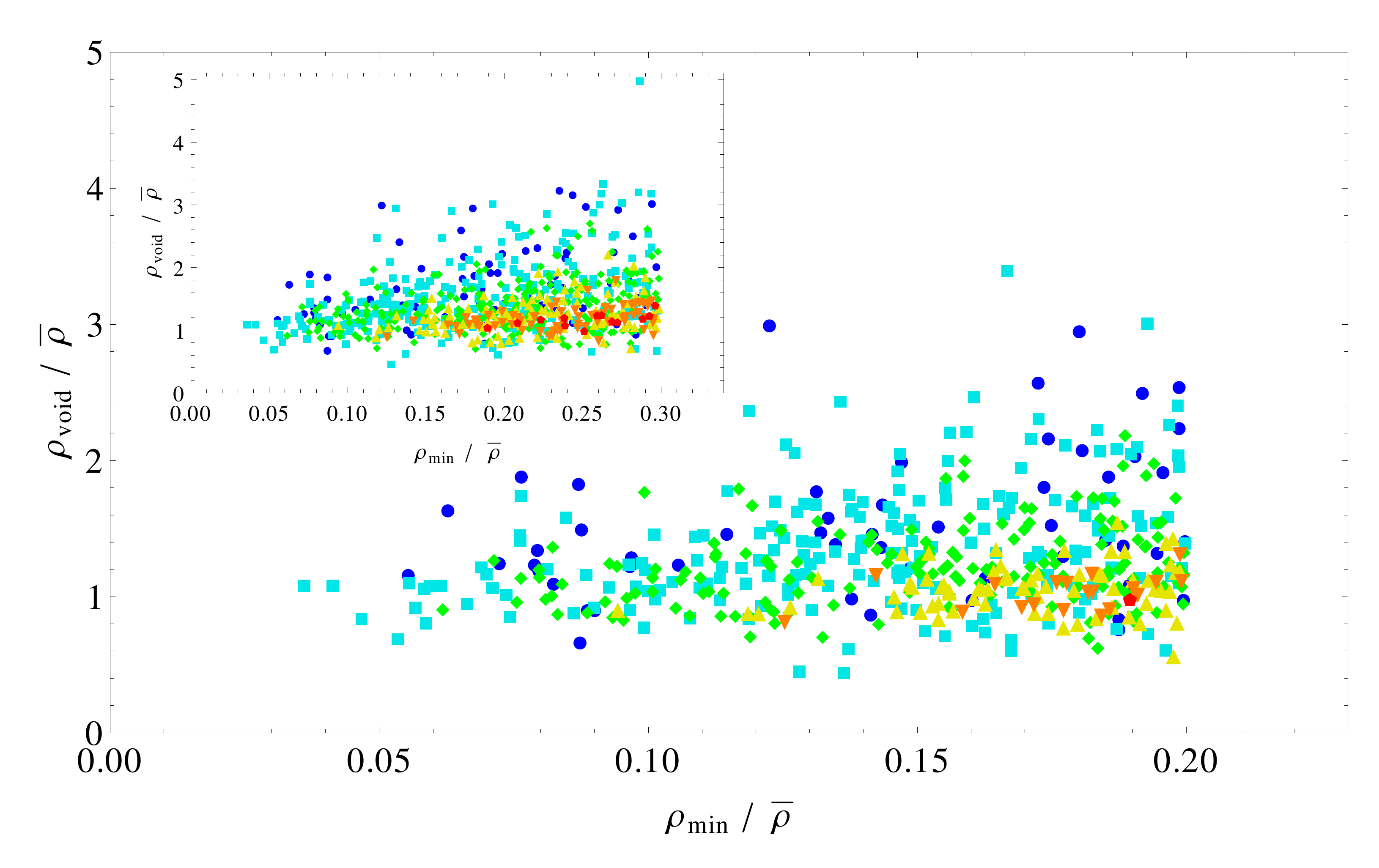}
\caption{Minimum density $\rho_\rmn{min}$ and average density $\rho_\rmn{void}$ values for Type2 voids. Colours and symbols are as in Figure~\ref{figure:Sutterrho}.
\emph{Inset:} The same figure for Type1 voids. Type1 voids have a larger spread of $\rho_\rmn{void}$ values, but note that for both types, the nature of the watershed algorithm means that the typical average density is of the order of the mean.} 
\label{figure:rhominrho}
\end{figure*}

Figure~\ref{figure:clusradiusredshift} shows the same effective radius distribution for superclusters. A similar resolution effect can clearly be seen in the increase of the minimum size with the increasing redshift and decreasing mean density of the sample. As expected for overdense objects, superclusters are typically both smaller and more numerous than voids. However, most superclusters are still far larger than galaxy cluster scales. They should therefore not be regarded as bound objects like clusters but simply as large-scale peaks in the galaxy density field similar to those identified by \cite{Granett:2008ju}. 

\subsection{Void densities and radial profiles}
\label{subsection:densities}

Figure~\ref{figure:rhominrho} shows the distribution of the minimum and average density values for Type1 and Type2 voids, and can be directly compared with Figure~\ref{figure:Sutterrho}. Both $\rho_\rmn{min}$ and $\rho_\rmn{void}$ have been normalized in units of the mean density, taking into account the appropriate correction for the redshift dependence of the mean. At such low minimum densities, there is at best only a very weak correlation between the two values. 

It is clear that despite the strict upper limit for the minimum density of the void, $\rho_\rmn{void}$ values are both noisy and biased high: this is because the watershed algorithm will generally include within a void galaxies from high-density walls and filaments at its edges. This is consistent with the observation that the lower redshift samples, where the growth of clusters is more pronounced and the higher number density of galaxies helps better resolve overdense filaments, have the largest spread in $\rho_\rmn{void}$. Also notice that the $\rho_\rmn{void}$ distribution is noisier for Type1 voids, which have looser criteria for the merging of zones.

In order to investigate the nature of the voids found and the applicability of the effective radius description, we reconstruct spherically averaged radial profiles of the stacked galaxy density about the void barycentres. To do this, we rescale all voids in a given sample by their effective radius, stack them so that their barycentres coincide, and count the number of galaxies contained within thin spherical shells about the barycentre. This number is then normalized in units of the expected number in a uniform distribution at the mean density for the sample as a whole, with no correction for the redshift dependence of this mean applied. In counting galaxies we also do not restrict ourselves to those identified as members of the void, but include \emph{all} galaxies within the radius limits. These simplifications mean that in effect we discard all the information about void shapes and the detailed topology of the density field, and simply model each void as a sphere of radius $R_\rmn{eff}$.

\begin{figure*}
\includegraphics[width=160mm]{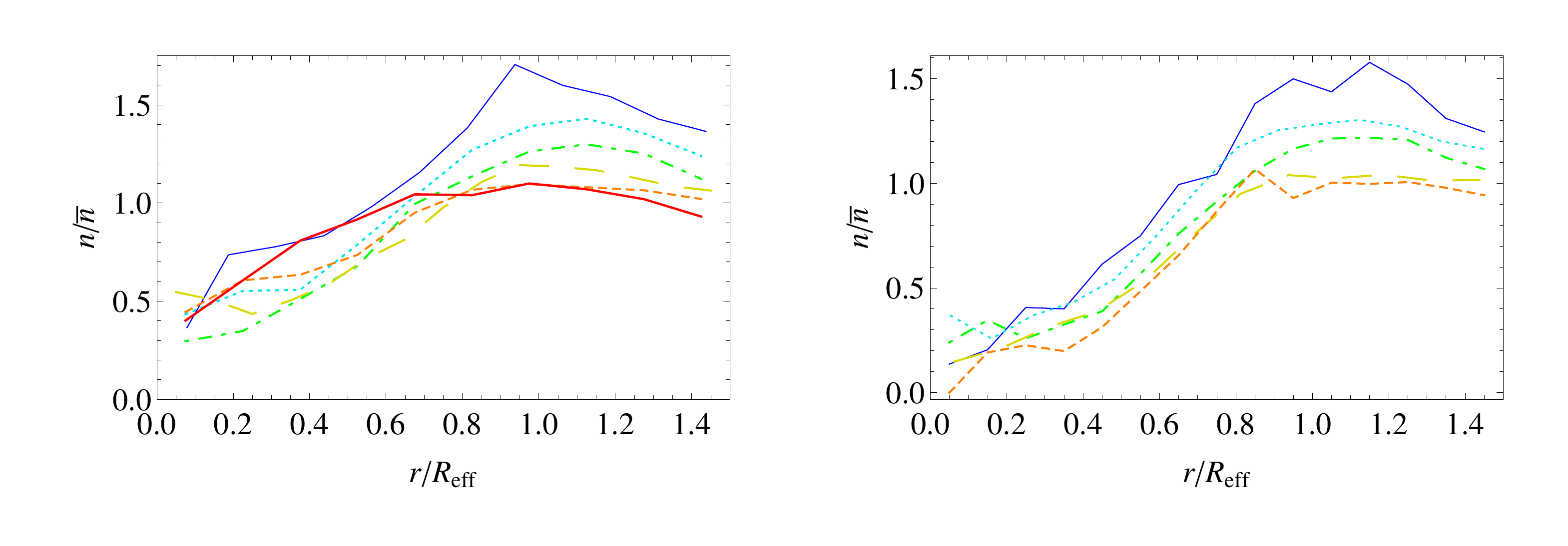}
\caption{Spherically averaged stacked radial profiles of voids. We rescale all voids by their effective radius $R_\rmn{eff}$ and count the number of all galaxies (not only void member galaxies) in spherical shells centred on the void barycentre. The counts are normalized in units of the sample mean number density, with no correction for redshift-dependent selection function.  \emph{Left panel:} Type1 void profiles. Voids from different samples are indicated as follows: \emph{dim1} (blue, thin solid line), \emph{dim2} (cyan, dotted), \emph{bright1} (green, dot-dashed), \emph{bright2} (yellow, long dashed), \emph{lrgdim} (orange, short dashed), \emph{lrgbright} (red, thick solid). \emph{Right panel:} The same for Type2 voids, but with \emph{lrgbright} omitted as it contains only one void.} 
\label{figure:profiles}
\end{figure*}

The radial profiles reconstructed in this manner are shown in Figure~\ref{figure:profiles}. As expected, the crude spherical model results in some noise in the profiles. Nevertheless, the same universal qualitative features can be observed for voids in all samples: an underdensity at the centre, with the density increasing towards a peak at $r\sim R_\rmn{eff}$ forming a lip. The height of the overdense peak at the lip increases for the more highly clustered and denser samples at lower redshift, whereas it is not very pronounced for \emph{lrgdim} and \emph{lrgbright}. This results in an overcompensated average profile for voids from the lower redshift samples, in accord with Figure~\ref{figure:rhominrho}. The central underdensities are deeper for Type2 voids than for Type1, as expected due to the tighter selection criteria on $\rho_\rmn{min}$. 

The most important conclusion that can be drawn from these stacked radial profiles is the confirmation that our methodology is successfully identifying locations of underdensities in the galaxy distribution, since the locations of the void barycentres are on average underdense. Another important result is confirmation that the simple characterization of voids by their $R_\rmn{eff}$ values is indeed meaningful. 

\begin{figure*}
\includegraphics[width=160mm]{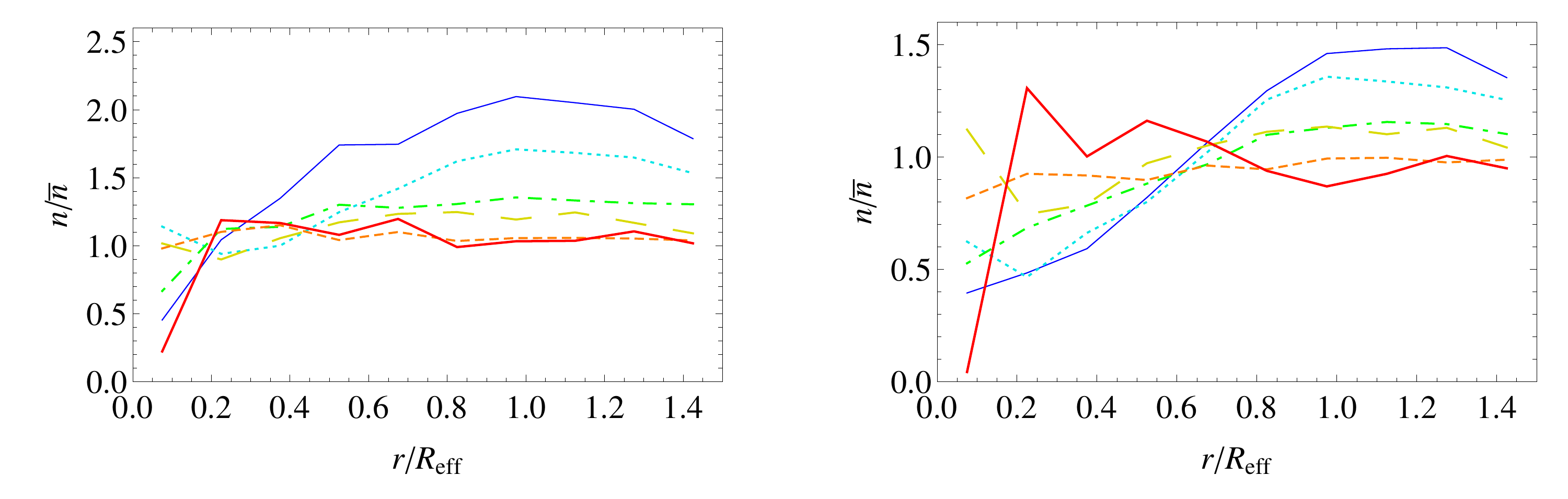}
\caption{Stacked radial profiles constructed as in Figure~\ref{figure:profiles} but for void locations and radii provided in the \citet{Sutter:2012wh} catalogue. Colours and line styles are the same as in Figure~\ref{figure:profiles}. \emph{Left panel:} Profiles when all listed voids are included in the stack. \emph{Right panel:} Profiles for stacks including only those voids which are listed as having central densities $\rho_\rmn{min}<0.3\overline{\rho}$ (see Section~\ref{section:Sutter}). Applying this additional cut excludes almost half the voids in the original catalogue, but only marginally improves the profiles for \emph{dim1} and \emph{dim2}, and not at all for the other samples.} 
\label{figure:Sutterprofiles}
\end{figure*}

For the sake of comparison, in Figure~\ref{figure:Sutterprofiles} we show the profiles reconstructed using exactly the same procedure with the claimed voids in the \cite{Sutter:2012wh} catalogue. Here there is no such universal behaviour, with the average density lying at or above the mean for almost all distances from the barycentre. Indeed stacked profiles from some samples fail to even show an underdensity at the centre. In this case modeling the claimed voids as spheres of the effective radius is meaningless, and their barycentre locations do not, on average, correspond to underdensities. This invalidates the use of this catalogue in several works \citep[e.g.][]{Sutter:2012tf, Pisani:2013yxa,Melchior:2013,Ilic:2013cn,Planck:ISW}, which assume the existence a universal radial profile based on the void effective radius. It could still be argued that the failure of the spherical model in this case does not strictly speaking preclude the existence of minima in the VTFE reconstructed density field at those locations, if the voids found by {\small ZOBOV} were not spherical on average. However, Figure~\ref{figure:profiles} shows that such a model works reasonably well for our catalogue, and only fails for the \cite{Sutter:2012wh} voids. 

It might be supposed that applying an additional quality cut on the quoted values of $\rho_\rmn{min}$ for the \citet{Sutter:2012wh} voids, similar to those employed for our Type1 and Type2 voids, would lead to a more robust catalogue (though the use of such an additional cut has neither been recommended nor implemented in the literature).\footnote{For reasons which are not explained in their paper, none of the \citet{Sutter:2012wh} voids consist of more than a single zone each. Therefore the additional criteria we employ on $\rho_\rmn{link}$ and $r$ are redundant in this case.} In fact, this is not the case. The right panel of Figure~\ref{figure:Sutterprofiles} shows the stacked average radial profiles after application of a cut $\rho_\rmn{min}<0.3\overline\rho$, which excludes almost half the listed voids. Here $\rho_\rmn{min}$ is determined from the data provided with the catalogue, as explained in Section~\ref{section:Sutter}. Although this somewhat improves the stacked profile for the \emph{dim1} and \emph{dim2} samples, on the whole the \citet{Sutter:2012wh} catalogue still fails to show any universal radial behaviour. Given the extent of the problems described in Section~\ref{section:Sutter}, this is not surprising.

While the simplified modeling of our voids as effective spheres is useful in some contexts, we caution against taking the profiles obtained above too seriously. Apart from the fact that most of the density information used by {\small ZOBOV} has been discarded, the very low mean galaxy number densities for the samples used also means that discreteness noise is an important problem if the radial bins are too narrow. The number of voids in some stacks is also small, particularly for the LRG samples. In addition, the number of voids that are further than a specified distance from a survey edge or hole in the mask falls as that distance increases. This can affect the determination of $R_\rmn{eff}$ due to truncation of the void volume, but also means that the stacked profile is somewhat uncertain at large radii $r>R_\rmn{eff}$ since part of the spherical ring about a given void may extend into a region not within the survey area, leading to an artificial decrease in the stacked density. We do not expect this problem to be very important over the limited radius range shown in Figure~\ref{figure:profiles}, but it may contribute to the slight downturn in densities seen at $r\simeq1.4R_\rmn{eff}$.



\section{Conclusions}
\label{section:conclusions}

We have used a modification of the void-finding algorithm {\small ZOBOV} to produce a catalogue of voids identified in several volume-limited and quasi-volume-limited samples of the SDSS DR7 main galaxy and LRG catalogues. The primary step in this process is the reconstruction of the density field using a Voronoi tessellation field estimator, which can present several complications. We handle the complex survey geometry and holes due to the bright star mask by using mock boundary particles to prevent voids from extending beyond the survey area. In addition, and in contrast to previous studies, we also use an appropriately conservative method for dealing with the contamination of the reconstructed density field by these mock particles. Our algorithm also includes a correction for the redshift-dependence of the mean number density of galaxies, which is particularly important for the LRG samples, and allows our method to be extended in the future to other datasets which may not be volume-limited.

The second step in the algorithm is to construct voids by joining basins around density minima according to the watershed transform. Although {\small ZOBOV} is often described as being parameter-free, in fact processing of the output after this step necessarily requires the introduction of one or more parameters to define what is meant by a void. We address this problem by introducing two different definitions of a void that are based on physical criteria that differentiate these structures from those that are artefacts of Poisson noise. Again this procedure is different from that followed in previous work. In the process, we clarify some misunderstandings about the operation of the watershed algorithm and which physical criteria are appropriate for use in void selection. We have discussed the differences in the properties of voids selected to match the two different definitions; however, as we realise that these are still somewhat arbitrary and that different definitions may prove to be useful in different contexts, we also provide in our catalogue information required for users to apply their own desired definitions according to need.

Our catalogue will be useful for several different studies of void properties, including the Alcock-Paczynski test, studies of the properties of galaxies within voids, gravitational lensing by voids, their imprints on the CMB through the integrated Sachs-Wolfe effect, and so on. For these purposes comparisons with the properties of void catalogues derived from simulated galaxy catalogues may be required for calibration and development of a theoretical understanding. In this respect the Jubilee Project simulation \citep{Watson:2013cxa} will be particularly important, as it provides a mock LRG catalogue as well as ISW and lensing information. We plan to investigate some of these issues in future work. 

Of course, another public catalogue of voids found in the same galaxy data by  \cite{Sutter:2012wh} using a different modification of the same algorithm has been available for some time, and has been used in a wide variety of studies. However, as we have shown in some detail, this catalogue suffers from several flaws in its construction. These include inadequate controls for boundary contamination, inappropriate choice of void selection criteria, and inconsistent application of even these criteria, resulting in the inclusion of many ``voids" that do not correspond to underdensities. We also have doubts about the physical interpretation of structures found in the redshift coordinates used for the primary catalogue provided. Given this lack of self-consistency, it is unsurprising that this catalogue is remarkably different to ours, as are the conclusions about void properties that may be drawn from it. In particular, Figure~\ref{figure:Sutterprofiles} demonstrates that the void locations listed in this catalogue do not, on average, correspond to underdense regions in the galaxy distribution. Our results suggest that previous studies that have used the \citet{Sutter:2012wh} catalogue may need to be reconsidered.

In addition to the void catalogue, we also provide a catalogue of ``superclusters", identified using the same watershed algorithm applied to the inverse of the reconstructed density field. With a few notable exceptions \citep{Granett:2008ju}, previous catalogues of superclusters \citep{Einasto:1997zd,Einasto:2003us,Berlind:2006,Einasto:2006mm,Einasto:2011zc,Liivamagi:2012} have identified superclusters using other methods, to which we believe the watershed algorithm is superior. Unlike galaxy clusters, superclusters probably do not form bound objects and so a theoretical description of their properties may prove to be difficult. However, as they are defined analogously to voids, they do correspond to the locations of large-scale peaks in the galaxy density, and so presumably also in the matter density field. Studies of their gravitational properties through lensing and ISW fluctuations may therefore prove illuminating.

All data products associated with this catalogue are available for public download at \url{www.hip.fi/nadathur/download/dr7catalogue}.

\section{Acknowledgements}
We thank Paul Sutter for correspondence and for comments on an earlier draft.
SN acknowledges support from the Sofja Kovalevskaja program of the Alexander von Humboldt Foundation, and from Academy of Finland grant 1263714. SH was supported by Academy of Finland grant 131454 and by the Science and Technology Facilities Council (grant number ST/I000976/1). 

This research has used data from the SDSS Data Release 7. Funding for the SDSS and SDSS-II has been provided by the Alfred P. Sloan Foundation, the Participating Institutions, the National Science Foundation, the U.S. Department of Energy, the National Aeronautics and Space Administration, the Japanese Monbukagakusho, the Max Planck Society, and the Higher Education Funding Council for England. The SDSS website is \url{http://www.sdss.org/}.

\bibliography{../../refs.bib}

\begin{thebibliography}{}

\bibitem[\protect\citeauthoryear{Abazajian et~al.,}{Abazajian
  et~al.}{2009}]{Abazajian:2008wr}
Abazajian K.~N.,  et~al., 2009, \apjs, 182, 543

\bibitem[\protect\citeauthoryear{{Achitouv}, {Neyrinck} \&
  {Paranjape}}{{Achitouv} et~al.}{2013}]{Achitouv:2013}
{Achitouv} I.,  {Neyrinck} M.,    {Paranjape} A.,  2013, ArXiv e-prints,
  1309.3799

\bibitem[\protect\citeauthoryear{Adelman-McCarthy et~al.,}{Adelman-McCarthy
  et~al.}{2008}]{AdelmanMcCarthy:2007aa}
Adelman-McCarthy J.~K.,  et~al., 2008, \apjs, 175, 297

\bibitem[\protect\citeauthoryear{{Alcock} \& {Paczynski}}{{Alcock} \&
  {Paczynski}}{1979}]{Alcock:1979}
{Alcock} C.,  {Paczynski} B.,  1979, \nat, 281, 358

\bibitem[\protect\citeauthoryear{{Berlind} et~al.,}{{Berlind}
  et~al.}{2006}]{Berlind:2006}
{Berlind} A.~A.  et~al., 2006, \apjs, 167, 1

\bibitem[\protect\citeauthoryear{Bertschinger}{Bertschinger}{1985}]{Bertsching%
er:1985nj}
Bertschinger E.,  1985, \apjs, 58, 1

\bibitem[\protect\citeauthoryear{{Biswas}, {Alizadeh} \& {Wandelt}}{{Biswas}
  et~al.}{2010}]{Biswas:2010}
{Biswas} R.,  {Alizadeh} E.,    {Wandelt} B.~D.,  2010, \prd, 82, 023002

\bibitem[\protect\citeauthoryear{{Blanton} et~al.,}{{Blanton}
  et~al.}{2005}]{Blanton:2004aa}
{Blanton} M.~R.  et~al., 2005, \aj, 129, 2562

\bibitem[\protect\citeauthoryear{Brunino, Trujillo, Pearce \& Thomas}{Brunino
  et~al.}{2007}]{Brunino:2006ym}
Brunino R.,  Trujillo I.,  Pearce F.~R.,    Thomas P.~A.,  2007, \mnras, 375,
  184

\bibitem[\protect\citeauthoryear{{Colberg} et~al.,}{{Colberg}
  et~al.}{2008}]{Colberg:2008}
{Colberg} J.~M.  et~al., 2008, \mnras, 387, 933

\bibitem[\protect\citeauthoryear{{D'Amico}, {Musso}, {Nore{\~n}a} \&
  {Paranjape}}{{D'Amico} et~al.}{2011}]{D'Amico:2010kh}
{D'Amico} G.,  {Musso} M.,  {Nore{\~n}a} J.,    {Paranjape} A.,  2011, \prd,
  83, 023521

\bibitem[\protect\citeauthoryear{Dubinski, Nicolaci~da Costa, Goldwirth, Lecar
  \& Piran}{Dubinski et~al.}{1993}]{Dubinski:1992tr}
Dubinski J.,  Nicolaci~da Costa L.,  Goldwirth D.,  Lecar M.,    Piran T.,
  1993, \apj, 410, 458

\bibitem[\protect\citeauthoryear{Einasto, Einasto, Hutsi, Saar, Tucker
  et~al.,}{Einasto et~al.}{2003}]{Einasto:2003us}
Einasto J.,  Einasto M.,  Hutsi G.,  Saar E.,  Tucker D.,    et~al., 2003,
  \aap, 410, 425

\bibitem[\protect\citeauthoryear{{Einasto} et~al.,}{{Einasto}
  et~al.}{2007}]{Einasto:2006mm}
{Einasto} J.  et~al., 2007, \aap, 462, 811

\bibitem[\protect\citeauthoryear{{Einasto}, {Liivam{\"a}gi}, {Tago}, {Saar},
  {Tempel}, {Einasto}, {Mart{\'{\i}}nez} \& {Hein{\"a}m{\"a}ki}}{{Einasto}
  et~al.}{2011}]{Einasto:2011zc}
{Einasto} M.,  {Liivam{\"a}gi} L.~J.,  {Tago} E.,  {Saar} E.,  {Tempel} E.,
  {Einasto} J.,  {Mart{\'{\i}}nez} V.~J.,    {Hein{\"a}m{\"a}ki} P.,  2011,
  \aap, 532, A5

\bibitem[\protect\citeauthoryear{{Einasto}, {Tago}, {Jaaniste}, {Einasto} \&
  {Andernach}}{{Einasto} et~al.}{1997}]{Einasto:1997zd}
{Einasto} M.,  {Tago} E.,  {Jaaniste} J.,  {Einasto} J.,    {Andernach} H.,
  1997, \aaps, 123, 119

\bibitem[\protect\citeauthoryear{Eisenstein et~al.,}{Eisenstein
  et~al.}{2001}]{Eisenstein:2001cq}
Eisenstein D.~J.,  et~al., 2001, \aj, 122, 2267

\bibitem[\protect\citeauthoryear{Fillmore \& Goldreich}{Fillmore \&
  Goldreich}{1984}]{Fillmore:1984wk}
Fillmore J.,  Goldreich P.,  1984, \apj, 281, 1

\bibitem[\protect\citeauthoryear{Flender, Hotchkiss \& Nadathur}{Flender
  et~al.}{2013}]{Flender:2012wu}
Flender S.,  Hotchkiss S.,    Nadathur S.,  2013, JCAP, 1302, 013

\bibitem[\protect\citeauthoryear{Forero-Romero, Hoffman, Gottloeber, Klypin \&
  Yepes}{Forero-Romero et~al.}{2009}]{ForeroRomero:2008ig}
Forero-Romero J.,  Hoffman Y.,  Gottloeber S.,  Klypin A.,    Yepes G.,  2009,
  \mnras, 396, 1815

\bibitem[\protect\citeauthoryear{Foster \& Nelson}{Foster \&
  Nelson}{2009}]{Foster:2009rt}
Foster C.,  Nelson L.~A.,  2009, \aj, 699, 1252

\bibitem[\protect\citeauthoryear{Furlanetto \& Piran}{Furlanetto \&
  Piran}{2006}]{Furlanetto:2005cc}
Furlanetto S.,  Piran T.,  2006, \mnras, 366, 467

\bibitem[\protect\citeauthoryear{{Goldberg} \& {Vogeley}}{{Goldberg} \&
  {Vogeley}}{2004}]{Goldberg:2004}
{Goldberg} D.~M.,  {Vogeley} M.~S.,  2004, \apj, 605, 1

\bibitem[\protect\citeauthoryear{Gorski, Hivon, Banday, Wandelt, Hansen
  et~al.,}{Gorski et~al.}{2005}]{Gorski:2004by}
Gorski K.,  Hivon E.,  Banday A.,  Wandelt B.,  Hansen F.,    et~al., 2005,
  \apj, 622, 759

\bibitem[\protect\citeauthoryear{Granett, Neyrinck \& Szapudi}{Granett
  et~al.}{2008}]{Granett:2008ju}
Granett B.~R.,  Neyrinck M.~C.,    Szapudi I.,  2008, \apj, 683, L99

\bibitem[\protect\citeauthoryear{Granett, Neyrinck \& Szapudi}{Granett
  et~al.}{2009}]{Granett:2008dz}
Granett B.~R.,  Neyrinck M.~C.,    Szapudi I.,  2009, \apj, 701, 414

\bibitem[\protect\citeauthoryear{{Gregory} \& {Thompson}}{{Gregory} \&
  {Thompson}}{1978}]{Gregory:1978}
{Gregory} S.~A.,  {Thompson} L.~A.,  1978, \apj, 222, 784

\bibitem[\protect\citeauthoryear{{Hamaus}, {Wandelt}, {Sutter}, {Lavaux} \&
  {Warren}}{{Hamaus} et~al.}{2013}]{Hamaus:2013}
{Hamaus} N.,  {Wandelt} B.~D.,  {Sutter} P.~M.,  {Lavaux} G.,    {Warren}
  M.~S.,  2013, ArXiv e-prints, 1307.2571

\bibitem[\protect\citeauthoryear{{Hern{\'a}ndez-Monteagudo} \&
  {Smith}}{{Hern{\'a}ndez-Monteagudo} \&
  {Smith}}{2013}]{HernandezMonteagudo:2012ms}
{Hern{\'a}ndez-Monteagudo} C.,  {Smith} R.~E.,  2013, \mnras, 435, 1094

\bibitem[\protect\citeauthoryear{Hoyle \& Vogeley}{Hoyle \&
  Vogeley}{2002}]{Hoyle:2001kn}
Hoyle F.,  Vogeley M.~S.,  2002, \apj, 566, 641

\bibitem[\protect\citeauthoryear{Hoyle \& Vogeley}{Hoyle \&
  Vogeley}{2004}]{Hoyle:2003hc}
Hoyle F.,  Vogeley M.~S.,  2004, \aj, 607, 751

\bibitem[\protect\citeauthoryear{Hunt \& Sarkar}{Hunt \&
  Sarkar}{2010}]{Hunt:2008wp}
Hunt P.,  Sarkar S.,  2010, \mnras, 401, 547

\bibitem[\protect\citeauthoryear{{Ili{\'c}}, {Langer} \& {Douspis}}{{Ili{\'c}}
  et~al.}{2013}]{Ilic:2013cn}
{Ili{\'c}} S.,  {Langer} M.,    {Douspis} M.,  2013, \aap, 556, A51

\bibitem[\protect\citeauthoryear{{J{\~o}eveer}, {Einasto} \&
  {Tago}}{{J{\~o}eveer} et~al.}{1978}]{Joeveer:1978}
{J{\~o}eveer} M.,  {Einasto} J.,    {Tago} E.,  1978, \mnras, 185, 357

\bibitem[\protect\citeauthoryear{{Jennings}, {Li} \& {Hu}}{{Jennings}
  et~al.}{2013}]{Jennings:2013nsa}
{Jennings} E.,  {Li} Y.,    {Hu} W.,  2013, \mnras, 434, 2167

\bibitem[\protect\citeauthoryear{{Kamionkowski}, {Verde} \&
  {Jimenez}}{{Kamionkowski} et~al.}{2009}]{Kamionkowski:2009}
{Kamionkowski} M.,  {Verde} L.,    {Jimenez} R.,  2009, \jcap, 1, 10

\bibitem[\protect\citeauthoryear{{Kazin} et~al.,}{{Kazin}
  et~al.}{2010}]{Kazin:2010}
{Kazin} E.~A.  et~al., 2010, \apj, 710, 1444

\bibitem[\protect\citeauthoryear{Kirshner, Oemler A., Schechter \&
  Shectman}{Kirshner et~al.}{1981}]{Kirshner:1981wz}
Kirshner R.,  Oemler A. J.,  Schechter P.,    Shectman S.,  1981, \apj, 248,
  L57

\bibitem[\protect\citeauthoryear{Komatsu et~al.,}{Komatsu
  et~al.}{2011}]{Komatsu:2010fb}
Komatsu E.,  et~al., 2011, \apjs, 192, 18

\bibitem[\protect\citeauthoryear{{Krause}, {Chang}, {Dor{\'e}} \&
  {Umetsu}}{{Krause} et~al.}{2013}]{Krause:2013}
{Krause} E.,  {Chang} T.-C.,  {Dor{\'e}} O.,    {Umetsu} K.,  2013, \apjl, 762,
  L20

\bibitem[\protect\citeauthoryear{Lavaux \& Wandelt}{Lavaux \&
  Wandelt}{2010}]{Lavaux:2009wm}
Lavaux G.,  Wandelt B.~D.,  2010, \mnras, 403, 1392

\bibitem[\protect\citeauthoryear{Lavaux \& Wandelt}{Lavaux \&
  Wandelt}{2012}]{Lavaux:2011yh}
Lavaux G.,  Wandelt B.~D.,  2012, \apj, 754, 109

\bibitem[\protect\citeauthoryear{Lee \& Park}{Lee \& Park}{2006}]{Lee:2006gj}
Lee J.,  Park D.,  2006, \apj, 652, 1

\bibitem[\protect\citeauthoryear{Lee \& Park}{Lee \& Park}{2009}]{Lee:2007kq}
Lee J.,  Park D.,  2009, \apj, 696, L10

\bibitem[\protect\citeauthoryear{Li, Zhao \& Koyama}{Li
  et~al.}{2012}]{Li:2011pj}
Li B.,  Zhao G.-B.,    Koyama K.,  2012, \mnras, 421, 3481

\bibitem[\protect\citeauthoryear{{Li} \& {Zhao}}{{Li} \&
  {Zhao}}{2009}]{Li:2009}
{Li} B.,  {Zhao} H.,  2009, \prd, 80, 044027

\bibitem[\protect\citeauthoryear{{Liivam{\"a}gi}, {Tempel} \&
  {Saar}}{{Liivam{\"a}gi} et~al.}{2012}]{Liivamagi:2012}
{Liivam{\"a}gi} L.~J.,  {Tempel} E.,    {Saar} E.,  2012, \aap, 539, A80

\bibitem[\protect\citeauthoryear{{Melchior}, {Sutter}, {Sheldon}, {Krause} \&
  {Wandelt}}{{Melchior} et~al.}{2013}]{Melchior:2013}
{Melchior} P.,  {Sutter} P.~M.,  {Sheldon} E.~S.,  {Krause} E.,    {Wandelt}
  B.~D.,  2013, ArXiv e-prints, 1309.2045

\bibitem[\protect\citeauthoryear{{Nadathur}}{{Nadathur}}{2013}]{Nadathur:2013m%
va}
{Nadathur} S.,  2013, \mnras, 434, 398

\bibitem[\protect\citeauthoryear{{Nadathur} \& {Hotchkiss}}{{Nadathur} \&
  {Hotchkiss}}{2013}]{NH:2013b}
{Nadathur} S.,  {Hotchkiss} S.,  2013, ArXiv e-prints, 1310.6911

\bibitem[\protect\citeauthoryear{Nadathur, Hotchkiss \& Sarkar}{Nadathur
  et~al.}{2012}]{Nadathur:2011iu}
Nadathur S.,  Hotchkiss S.,    Sarkar S.,  2012, JCAP, 1206, 042

\bibitem[\protect\citeauthoryear{{Neyrinck}}{{Neyrinck}}{2008}]{Neyrinck:2007g%
y}
{Neyrinck} M.~C.,  2008, \mnras, 386, 2101

\bibitem[\protect\citeauthoryear{Neyrinck, Gnedin \& Hamilton}{Neyrinck
  et~al.}{2005}]{Neyrinck:2004gj}
Neyrinck M.~C.,  Gnedin N.~Y.,    Hamilton A.~J.,  2005, \mnras, 356, 1222

\bibitem[\protect\citeauthoryear{Padmanabhan, Schlegel, Finkbeiner, Barentine,
  Blanton et~al.,}{Padmanabhan et~al.}{2008}]{Padmanabhan:2007zd}
Padmanabhan N.,  Schlegel D.,  Finkbeiner D.,  Barentine J.,  Blanton M.,
  et~al., 2008, \apj, 674, 1217

\bibitem[\protect\citeauthoryear{Pan, Vogeley, Hoyle, Choi \& Park}{Pan
  et~al.}{2012}]{Pan:2011hx}
Pan D.~C.,  Vogeley M.~S.,  Hoyle F.,  Choi Y.-Y.,    Park C.,  2012, \mnras,
  421, 926

\bibitem[\protect\citeauthoryear{Paranjape, Lam \& Sheth}{Paranjape
  et~al.}{2012}]{Paranjape:2011bz}
Paranjape A.,  Lam T.~Y.,    Sheth R.~K.,  2012, \mnras, 420, 1648

\bibitem[\protect\citeauthoryear{{Park}, {Choi}, {Kim}, {Gott} III, {Kim} \&
  {Kim}}{{Park} et~al.}{2012}]{Park:2012dn}
{Park} C.,  {Choi} Y.-Y.,  {Kim} J.,  {Gott} III J.~R.,  {Kim} S.~S.,    {Kim}
  K.-S.,  2012, \apj, 759, L7

\bibitem[\protect\citeauthoryear{{Park} \& {Lee}}{{Park} \&
  {Lee}}{2007}]{Park:2007}
{Park} D.,  {Lee} J.,  2007, \apj, 665, 96

\bibitem[\protect\citeauthoryear{{Pisani}, {Lavaux}, {Sutter} \&
  {Wandelt}}{{Pisani} et~al.}{2013}]{Pisani:2013yxa}
{Pisani} A.,  {Lavaux} G.,  {Sutter} P.~M.,    {Wandelt} B.~D.,  2013, ArXiv
  e-prints, 1306.3052

\bibitem[\protect\citeauthoryear{{Planck Collaboration} et~al.,}{{Planck
  Collaboration} et~al.}{2013}]{Planck:ISW}
{Planck Collaboration} et~al., 2013, ArXiv e-prints, 1303.5079

\bibitem[\protect\citeauthoryear{Platen, van~de Weygaert \& Jones}{Platen
  et~al.}{2007}]{Platen:2007qk}
Platen E.,  van~de Weygaert R.,    Jones B.~J.,  2007, \mnras, 380, 551

\bibitem[\protect\citeauthoryear{{Ryden}}{{Ryden}}{1995}]{Ryden:1995}
{Ryden} B.~S.,  1995, \apj, 452, 25

\bibitem[\protect\citeauthoryear{{Ryden} \& {Melott}}{{Ryden} \&
  {Melott}}{1996}]{Ryden:1996}
{Ryden} B.~S.,  {Melott} A.~L.,  1996, \apj, 470, 160

\bibitem[\protect\citeauthoryear{{Schaap}}{{Schaap}}{2007}]{Schaap:2007}
{Schaap} W.~E.,  2007, PhD Thesis, University of Gr\"oningen

\bibitem[\protect\citeauthoryear{Sheth \& van~de Weygaert}{Sheth \& van~de
  Weygaert}{2004}]{Sheth:2003py}
Sheth R.~K.,  van~de Weygaert R.,  2004, \mnras, 350, 517

\bibitem[\protect\citeauthoryear{Strauss et~al.,}{Strauss
  et~al.}{2002}]{Strauss:2002dj}
Strauss M.~A.,  et~al., 2002, \aj, 124, 1810

\bibitem[\protect\citeauthoryear{{Suto}, {Sato} \& {Sato}}{{Suto}
  et~al.}{1984}]{Suto:1984}
{Suto} Y.,  {Sato} K.,    {Sato} H.,  1984, Progress of Theoretical Physics,
  71, 938

\bibitem[\protect\citeauthoryear{Sutter, Lavaux, Wandelt \& Weinberg}{Sutter
  et~al.}{2012a}]{Sutter:2012wh}
Sutter P.,  Lavaux G.,  Wandelt B.~D.,    Weinberg D.~H.,  2012a, \apj, 761, 44

\bibitem[\protect\citeauthoryear{Sutter, Lavaux, Wandelt \& Weinberg}{Sutter
  et~al.}{2012b}]{Sutter:2012tf}
Sutter P.,  Lavaux G.,  Wandelt B.~D.,    Weinberg D.~H.,  2012b, \apj, 761,
  187

\bibitem[\protect\citeauthoryear{{Sutter}, {Lavaux}, {Wandelt}, {Hamaus},
  {Weinberg} \& {Warren}}{{Sutter} et~al.}{2013}]{Sutter:2013ssy}
{Sutter} P.~M.,  {Lavaux} G.,  {Wandelt} B.~D.,  {Hamaus} N.,  {Weinberg}
  D.~H.,    {Warren} M.~S.,  2013, ArXiv e-prints, 1309.5087

\bibitem[\protect\citeauthoryear{{Sutter}, {Lavaux}, {Wandelt}, {Weinberg} \&
  {Warren}}{{Sutter} et~al.}{2013}]{Sutter:2013DR9}
{Sutter} P.~M.,  {Lavaux} G.,  {Wandelt} B.~D.,  {Weinberg} D.~H.,    {Warren}
  M.~S.,  2013, ArXiv e-prints, 1310.7155

\bibitem[\protect\citeauthoryear{{Swanson}, {Tegmark}, {Hamilton} \&
  {Hill}}{{Swanson} et~al.}{2008}]{Swanson:2008}
{Swanson} M.~E.~C.,  {Tegmark} M.,  {Hamilton} A.~J.~S.,    {Hill} J.~C.,
  2008, \mnras, 387, 1391

\bibitem[\protect\citeauthoryear{{Tavasoli}, {Vasei} \& {Mohayaee}}{{Tavasoli}
  et~al.}{2013}]{Tavasoli:2013}
{Tavasoli} S.,  {Vasei} K.,    {Mohayaee} R.,  2013, \aap, 553, A15

\bibitem[\protect\citeauthoryear{{Viel}, {Colberg} \& {Kim}}{{Viel}
  et~al.}{2008}]{Viel:2008}
{Viel} M.,  {Colberg} J.~M.,    {Kim} T.-S.,  2008, \mnras, 386, 1285

\bibitem[\protect\citeauthoryear{{Watson} et~al.,}{{Watson}
  et~al.}{2013}]{Watson:2013cxa}
{Watson} W.~A.  et~al., 2013, ArXiv e-prints, 1307.1712

\end{thebibliography}
\bibliographystyle{mn2e}

\appendix
\section{Layout of the catalogue data}

\begin{table*}
\begin{minipage}{118mm}
\caption{Primary information about the Type1 voids found in the \emph{dim1} galaxy sample. The full table can be downloaded from \url{www.hip.fi/nadathur/download/dr7catalogue} and contains a total of 80 rows and 10 columns. Only a portion is shown here for guidance. }
\begin{tabular}{@{}ccccccccc}
\hline
Void ID & RA (deg) & Dec (deg) & $z$ & $R_\rmn{eff}\;(h^{-1}$Mpc) & $\theta_\rmn{eff}$ (deg) & $\rho_\rmn{void}$ & $\rho_\rmn{min}$ & $r$ \\
\hline
10875 & 187.86 & 23.25 & 0.036 &	33.12 & 17.48 & 1.153 & 0.056 & 3.29 \\
10099 & 174.94 & 35.78 & 0.019 & 20.19 & 20.53 &1.703 & 0.063 & 5.13\\
10886 & 165.30 & 20.89 & 0.039 &	17.01 & 8.49 & 1.238 & 0.072 & 2.54\\
10418 & 228.22 & 18.44 & 0.033 & 16.52 & 9.70 & 1.874 & 0.076 & 3.05\\
\hline\\
\end{tabular}
\label{table:sampleinfo}
\end{minipage}
\end{table*}


The entire catalogue is provided as a single downloadable \verb"gzip"-archived file available from \url{www.hip.fi/nadathur/download/dr7catalogue}. Two versions of the catalogue are provided: the primary version in comoving coordinates, and a secondary version in redshift coordinates as discussed above. These are separated in directories labelled \verb"comovcoords" and \verb"redshiftcoords". 

Each of these directories is further divided into six folders containing the Type1 and Type2 void catalogues and the supercluster catalogue for each of the galaxy samples analysed here, and a folder called \verb"tools", which contains data useful for users wishing to apply their own selection criteria. The basic information provided includes the location of the barycentre of each structure, its volume, effective radius, average density and minimum or maximum density, its core galaxy and seed zone, the total number of galaxies in the seed zone, the number of zones merged to form the structure, the total number of particles in the structure, and its density ratio. These are split between two files for each structure type and each sample, named \verb"xxx_info.txt" and \verb"xxx_list.txt", where \verb"xxx" refers to the structure type. It is also possible to extract lists of member galaxies of each structure and their magnitudes. An example {\small PYTHON} script, \verb"postproc.py", demonstrates how to access this information and how to build alternative catalogues using user-defined selection criteria.

Table~\ref{table:sampleinfo} provides a sample of the contents of the \verb"Type1voids_info.txt" file, for structures found in the \emph{dim1} galaxy sample.


\label{lastpage}
\end{document}